\documentclass[
    reprint,
    cha,
    aip,
    twocolumn,
    superscriptaddress
]{revtex4-2}

\usepackage{adjustbox}
\usepackage[utf8]{inputenc}
\usepackage{lmodern}

\usepackage[dvipsnames]{xcolor}
\usepackage{amsmath}
\usepackage{amssymb}
\usepackage{graphicx}
\usepackage{overpic}
\usepackage{subfigure}
\usepackage{hhline}
\usepackage{transparent}
\usepackage{tikz}

\usepackage{bm}
\SetSymbolFont{largesymbols}{bold}{OMX}{txex}{b}{n}

\usepackage{physics}

\usepackage{relsize}

\usepackage[colorlinks=true, linkcolor=blue, citecolor=blue, hypertexnames=false]{hyperref}
\hypersetup{colorlinks=true, urlcolor=blue}
\usepackage[all]{hypcap}

\usepackage{mathtools}
\DeclarePairedDelimiterX{\set}[1]{\{}{\}}{\setargs{#1}}
\NewDocumentCommand{\setargs}{>{\SplitArgument{1}{;}}m}
{\setargsaux#1}
\NewDocumentCommand{\setargsaux}{mm}
{\IfNoValueTF{#2}{#1} {#1\nonscript\:\delimsize\vert\allowbreak\nonscript\:\mathopen{}#2}}

\begin{document}

\title{Measuring Correlation and Entanglement between Molecular Orbitals on a Trapped-Ion Quantum Computer}

\author{Gabriel Greene-Diniz}\thanks{email: gabriel.greene-diniz@quantinuum.com}

\affiliation{Quantinuum, Terrington House, 13-15 Hills Road, Cambridge CB2 1NL, United Kingdom}

\author{Chris N. Self}\thanks{email: christopher.self@quantinuum.com}

\affiliation{Quantinuum, Partnership House, Carlisle Place, London SW1P 1BX, United Kingdom}

\author{Michal Krompiec}
\affiliation{Quantinuum, Terrington House, 13-15 Hills Road, Cambridge CB2 1NL, United Kingdom}

\author{Luuk Coopmans}
\affiliation{Quantinuum, Partnership House, Carlisle Place, London SW1P 1BX, United Kingdom}

\author{Marcello Benedetti}
\affiliation{Quantinuum, Partnership House, Carlisle Place, London SW1P 1BX, United Kingdom}

\author{David Mu{\~n}oz Ramo}
\affiliation{Quantinuum, Terrington House, 13-15 Hills Road, Cambridge CB2 1NL, United Kingdom}

\author{Matthias Rosenkranz}
\affiliation{Quantinuum, Partnership House, Carlisle Place, London SW1P 1BX, United Kingdom}

\date{August 6, 2025} 

\begin{abstract}
\textsf{Quantifying correlation and entanglement between molecular orbitals can elucidate the role of quantum effects in strongly correlated reaction processes. However, accurately storing the wavefunction for a classical computation of those quantities can be prohibitive. Here we use the Quantinuum H1-1 trapped-ion quantum computer to calculate von Neumann entropies which quantify the orbital correlation and entanglement in a strongly correlated molecular system relevant to lithium-ion batteries (vinylene carbonate interacting with an O$_2$ molecule). As shown in previous works, fermionic superselection rules decrease correlations and reduce measurement overheads for constructing orbital reduced density matrices. Taking into account superselection rules we further reduce the number of measurements by finding commuting sets of Pauli operators. Using low overhead noise reduction techniques, we calculate von Neumann entropies in excellent agreement with noiseless benchmarks, indicating that correlations and entanglement between molecular orbitals can be accurately estimated from a quantum computation. Our results show that the one-orbital entanglement vanishes unless opposite-spin open shell configurations are present in the wavefunction.}
\end{abstract}

\maketitle

\section*{Introduction}\label{sec:intro}

\noindent A number of works have appeared recently that use quantum information theory to gain new insights into quantum chemical phenomena~\cite{aliverti-piuri24, wu24, scholes23}. This includes the strength and nature of electronic correlation in molecules~\cite{ding23, materia24, mottet14}, the behavior of an electronic wavefunction during a reaction (e.g. dissociation) in the presence of quasidegeneracy and/or static correlation~\cite{gersdorf97, boguslawski13, knecht14, ding20}, the accuracy of a particular active space selection~\cite{boguslawski2012entanglement, stein16, ding23AS}, and a novel understanding of chemical bonds containing multiorbital correlations~\cite{szalay17}. 

Entanglement and correlation play a key role in the quantum information theory framework~\cite{boguslawski15, ding20}, making strong correlation in chemistry a natural topic for study.
Measures of correlation and entanglement can be obtained from the mutual information~\cite{szalay17, vedral97, henderson01, rissler06} via the appropriate orbital entropies~\cite{boguslawski13, stein2017measuring, ding20, ding2020concept, zhang21}, yielding interesting characterizations of bond-breaking processes, and allowing for quantification of quantum and classical correlation contributions to chemical bonds~\cite{ding2020concept}. 
However, it is important to carefully separate the different ways that correlation arises in the chemical system so that the extent of the quantum effects are not overestimated.
The question of how quantum or classical the observed correlations are has been addressed for several molecular systems~\cite{ding2020concept, ding23, materia24}. 
The basis set dependency of correlation measures has also been highlighted~\cite{szalay17}, with a recent work emphasizing the danger of overestimations in the absence of atomically localized orbitals~\cite{materia24}. 
In addition, concerns have been raised regarding the overestimation of entanglement due to lack of adherence to fundamental fermionic symmetries, known as superselection rules (SSRs)~\cite{ding20, ding22}, which in turn raises interesting questions about the role of symmetry in the simulation of strongly correlated electrons. 

Previously orbital correlation and entanglement have been calculated using classical computational resources. Here we demonstrate a procedure for obtaining orbital entropies and two-orbital mutual information on a quantum computer. 
Specifically, we reconstruct the orbital reduced density matrices (ORDMs)~\cite{rissler06, boguslawski13, ding23} from measurement circuits executed on the Quantinuum \mbox{H1-1} trapped-ion quantum computer~\cite{h11}. The use of a quantum computer in this work is motivated by two main factors: the potential of quantum hardware to store the chemical wavefunction more efficiently than classical hardware~\cite{peruzzo14, mcardle20}, and a low number of measurable circuits due to grouping of Pauli operators into commuting sets when accounting for SSRs. Hence, quantum hardware is utilized here to demonstrate the methodology and facilitate future studies on larger systems.

As a model system, we study the formation of tetraoxabicyclo[3.2.0]heptan-3-one (substituted dioxetane) from the precursors 1,3-dioxol-2-one (vinylene carbonate, abbreviated as VC) and singlet oxygen ($^1$O$_2$), which is the addition of singlet oxygen to the double bond to form a dioxetane ring. Henceforth we refer to the reaction product as ``dioxetane'', noting that it is a substituted derivative of C$_2$O$_2$H$_4$. Recent studies have shown that $^1$O$_2$ tends to attack the hydrocarbon group of carbonates~\cite{mullinax21}, which is relevant to the degradation of carbonate solvents in oxide batteries since $^1$O$_2$ is produced during their operation~\cite{mahne17}. This reaction also involves a transition state which exhibits strong static correlation, as oxygen approaches the hydrocarbon termination and bonds are stretched. 
We calculate orbital von Neumann entropies in order to study the correlation and entanglement between strongly correlated molecular orbitals that play a significant role in the VC + $^1$O$_2$ $\rightarrow$ dioxetane reaction. This set of molecular orbitals are constructed by first using the nudged elastic band (NEB) method to determine the atomic geometries then applying an atomic valence active space (AVAS)~\cite{avas} projection to the $p$ orbitals of the O$_2$, described in more detail in the \hyperlink{methods}{Methods} section. 

To prepare the ground state wavefunctions at different steps of the VC + $^1$O$_2$ $\rightarrow$ dioxetane reaction on a quantum computer we encode the fermionic problem into qubits using a Jordan-Wigner (JW) transformation and then offline optimize a variational quantum eigensolver (VQE) ansatz that prepares the relevant states. 
With these wavefunctions prepared by running the circuits, we obtain orbital von Neumann entropies from eigenvalues of the ORDM, which are estimated from measurements on the quantum hardware (for more details, see the \hyperlink{methods}{Methods} section). To deal with noise we apply a low overhead post-measurement noise reduction scheme to the measured ORDMs, involving a thresholding method to filter out small singular values from  the noisy ORDMs~\cite{donoho2013optimal}, followed by a maximum likelihood estimate to reconstruct the physical ORDMs~\cite{smolin2012efficient}.

Our results show that orbital von Neumann entropies can be reliably calculated for moderate system sizes on a trapped-ion quantum computer.
The resulting orbital entropies paint a reasonable picture of the transition state of the VC+$^1$O$_2$ system: 2$p$ O orbitals are strongly correlated as oxygen bonds are stretched to align to the C-C bond of the carbonate, followed by a settling to the weakly correlated ground state of dioxetane, as reflected in the orbital entropies gleaned from one and two orbital ORDMs (1-ORDM and 2-ORDM, respectively). 

Additionally, previous works have investigated the inclusion of fundamental fermionic symmetries for the correct quantification of orbital entanglement and correlation~\cite{ding20, ding2020concept, ding22}. Our work also adds to these considerations by demonstrating an interesting consequence of incorporating fermionic superselection rules~\cite{wick52, friis16, ding2020concept} while also partitioning the measurable Pauli operators into commuting sets when constructing the ORDMs. Specifically, these superselection rules lead to a significant reduction in the number of circuits that need to be measured when evaluating the ORDM elements. Finally, by investigating orbital correlation and entanglement for both singlet and triplet spin configurations of the molecular states, our results highlight an important consequence of quantifying one-orbital entanglement via the 1-ORDM in a molecular orbital basis, namely that the absence of open shell spin configurations leads to vanishing one-orbital entanglement when superselection rules are taken into account. 

\hypertarget{methods}{\section*{Methods}}

\hypertarget{cchem}{\subsection*{Classical Computational Chemistry}}\label{sec:methods:cchem}

\begin{figure*}[ht]
    \centering
    \includegraphics[width=0.92\textwidth]{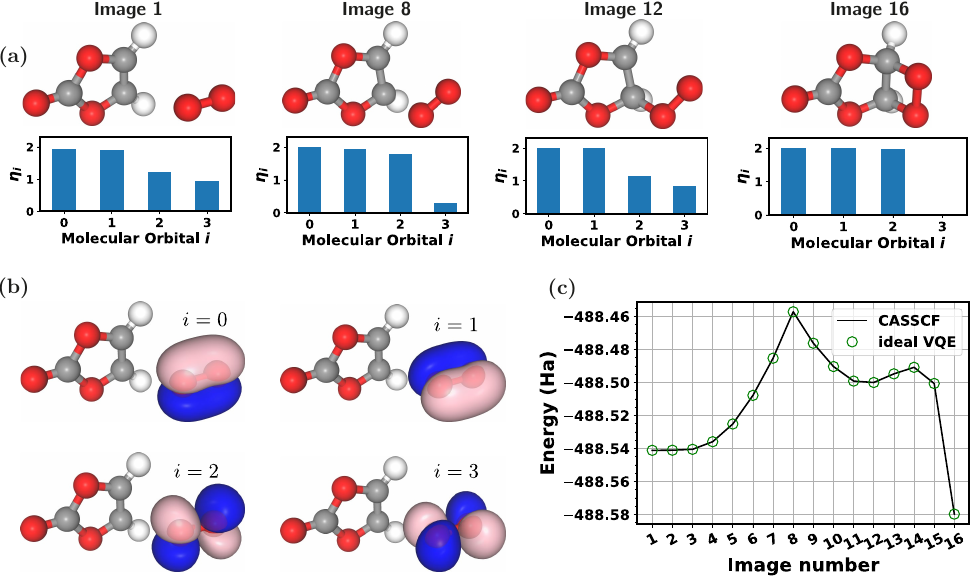}\hspace{0.01\textwidth}
    \caption{Molecular structures, orbitals, and total energies, of the VC + $^1$O$_2$ $\rightarrow$ dioxetane reaction, in which the singlet symmetry constraint is imposed on CASSCF calculations.
    (a) Four snapshots of the 16 NEB images generated for the reaction path. Initial geometry (image 1), intermediate transition states (image 8, image 12), and final geometry corresponding to dioxetane (image 16), of the NEB reaction path are shown. Oxygen atoms are red, carbon atoms are gray, and hydrogen atoms are white. Bar charts of natural orbital populations $\eta_{i}$ shown underneath NEB snapshots.
    (b) Contour plots of four orbitals corresponding to the selected active space (6 electrons in 4 orbitals) within the AVAS set constructed by projecting the oxygen $p$ orbitals of the O$_2$ molecule. These represent image 1 of the NEB path. Orbital indexes $i$ also shown. (c) Energies obtained from CASSCF calculations and compared to VQE, for all 16 images of the NEB reaction path representing VC + $^1$O$_2$ $\rightarrow$ dioxetane, using the (4, 6) subset of the AVAS set.
    }
\label{fig:neb_avas_cas}
\end{figure*}

\begin{table*}[t]
    \centering
    \noindent\adjustbox{max width=\textwidth}{
    \begin{tabular}{|c|c|c|c|c|}
        \hline
        NEB image & Quantum chemical statevector &
        PhasedX & $R_Z$ & ZZPhase \\[5pt]
        \hline
        1 & $\begin{array} {lcl} \ \ - 0.7246799 \boldsymbol | \ (11)_0, (11)_1, (11)_2, (00)_3 \ \boldsymbol \rangle + 0.6249013 \boldsymbol | \ (11)_0, (11)_1, (00)_2, (11)_3 \ \boldsymbol \rangle \\+ \ 0.2215456 \boldsymbol | \ (11)_0, (00)_1, (11)_2, (11)_3 \ \boldsymbol \rangle - 0.1877628 \boldsymbol | \ (00)_0, (11)_1, (11)_2, (11)_3 \ \boldsymbol \rangle\end{array}$ & 22 & 5 & 11 \\[10pt]
        \hline
        8 & $\begin{array} {lcl} \ \ \ 0.9268118 \boldsymbol | \ (11)_0, (11)_1, (11)_2, (00)_3 \ \boldsymbol \rangle - 0.3281654 \boldsymbol | \ (11)_0, (11)_1, (00)_2, (11)_3 \ \boldsymbol \rangle \\- \ 0.1771431 \boldsymbol| \ (11)_0, (00)_1, (11)_2, (11)_3 \ \boldsymbol \rangle + 0.0441317 \boldsymbol | \ (00)_0, (11)_1, (11)_2, (11)_3 \ \boldsymbol \rangle\end{array}$ & 22 & 4 & 11 \\[10pt]
        \hline
        12 & $\begin{array} {lcl} \ \ -0.7575021 \boldsymbol | \ (11)_0, (11)_1, (11)_2, (00)_3 \ \boldsymbol \rangle + 0.6526460 \boldsymbol | \ (11)_0, (11)_1, (00)_2, (11)_3 \ \boldsymbol \rangle \\+ \ 0.0126794 \boldsymbol | \ (11)_0, (00)_1, (11)_2, (11)_3 \ \boldsymbol \rangle - 0.0091089 \boldsymbol | \ (00)_0, (11)_1, (11)_2, (11)_3 \ \boldsymbol \rangle\end{array}$ & 22 & 5 & 11 \\[10pt]
        \hline
        16 & $\begin{array} {lcl} \ \ \ 0.9937562 \boldsymbol | \ (11)_0, (11)_1, (11)_2, (00)_3 \ \boldsymbol \rangle - 0.1098397 \boldsymbol | \ (11)_0, (11)_1, (00)_2, (11)_3 \ \boldsymbol \rangle \\- \ 0.0142564 \boldsymbol | \ (11)_0, (00)_1, (11)_2, (11)_3 \ \boldsymbol \rangle - 0.0134371 \boldsymbol | \ (00)_0, (11)_1, (11)_2, (11)_3 \ \boldsymbol \rangle\end{array}$ & 22 & 4 & 11
        \\[9pt]
        \hline
    \end{tabular}
    }
    \caption{Basis states of the VQE optimized statevector of 4 NEB images along the reaction path, in which the CASSCF optimizations were constrained to produce the $\langle S^2 \rangle = 0$ singlet. The orbital basis is arranged such that all basis states have the form $|\dots, (n^{\uparrow} \ n^{\downarrow})_i, (n^{\uparrow} \ n^{\downarrow})_{i+1}, \dots \rangle $, where $i$ labels the molecular orbitals (shown in Fig.~\ref{fig:neb_avas_cas} for image 1), and $n^{\sigma}$ is the $\sigma$ spin orbital occupation which corresponds to a qubit state (with qubit index $2i$ ($2i+1$) for $\sigma=\ \uparrow \ (\downarrow)$) after JW transformation. The basis states were initially chosen from the largest weight Slater determinants in CASSCF  and their coefficients were then optimized in VQE to produce (ideal, noiseless) energies that match the CASSCF energies to less than 10$^{-6}$ Hartrees. State circuits prepared using an adaption of a methodology involving controlled Givens rotations~\cite{arrazola22}, and circuits compiled to the Quantinuum H1-1 device. Rightmost columns report the number of one-qubit PhasedX and $R_Z$ gates, and two-qubit ZZPhase gates (see Appendix C of the Supplementary Information for more details).}
    \label{tab:sv_coeffs_gates}
\end{table*}

\noindent Initially, the minimum-energy path of the \mbox{VC + $^1$O$_2$ $\rightarrow$ dioxetane} reaction is determined with the nudged elastic band (NEB) method, using PySCF and Jónsson's group implementation of NEB~\cite{KNARR} available in the ASH package~\cite{ASH}. In the NEB calculations energies are computed using Density Functional Theory (DFT), approximated with the PBE exchange-correlation functional~\cite{pbe}. Throughout this work the atomic basis set used is def2-SVP~\cite{weigend05, basis_set}. For DFT calculations, finite temperature electronic smearing is used to handle the quasi-degeneracy of orbitals during the self-consistent field optimization. Four examples of the resulting geometries (or NEB ``images'') are shown in Fig.~\ref{fig:neb_avas_cas}a, and 16 images in total are determined. 

Wavefunction-based methods are subsequently applied to the geometries extracted from the NEB calculations, using the PySCF package~\cite{pyscf}. To this end, AVAS~\cite{avas} projections are carried out to narrow down the active space most relevant to the static correlation of the reaction. As an additional benefit, since AVAS involves projection onto targeted atomic orbitals~\cite{avas}, an intrinsically localized orbital basis is obtained which can help avoid the overestimation of correlation from more disperse orbital bases (as evidenced by orbitals that diagonalize the one-electron reduced density matrix compared to canonical molecular orbitals~\cite{materia24}), and consistent with the study of correlation among (atomically) localized subsystems~\cite{szalay17}. The AVAS method (for more details, we refer the reader to~\cite{avas}.) requires a particular choice of local atomic orbitals against which to project the canonical orbitals. This choice can be aided by chemical intuition: Since the O$_2$ molecule stretches and hybridizes with the carbons of VC, it stands to reason that a significant portion of the strong correlation is localized to the oxygen $p$ orbitals of the O$_2$ molecule. Hence the latter are chosen for the atomic orbital projections of AVAS. The resulting AVAS set yields 6 electrons in 9 molecular orbitals.

In order to make subsequent quantum computations more computationally economical, a subset of this AVAS set corresponding to the 4 energetically shallowest molecular orbitals (labelled as (4, 6), i.e. 6 electrons distributed among 4 molecular orbitals) is selected from the larger set of 9 molecular orbitals. These are used as the initial orbitals in complete active space self consistent field (CASSCF)~\cite{pyscf, szabo_ostlund} calculations, which correspond to  optimizing (\textit{i}) the coefficients of all symmetry-preserving basis states (Slater determinants) within the active space of orbitals, and (\textit{ii}) the coefficients of those active molecular orbitals. In this work, CASSCF is used to determine the most important electronic configurations contributing to the chemical wavefunction, which guide the state preparation as described in the \hyperlink{qchem}{Quantum Computing Chemical States} subsection. The total energies obtained from CASSCF along the NEB path are plotted in Fig.~\ref{fig:neb_avas_cas}c. These calculations are performed by imposing a constraint on the total spin operator $\langle S^2 \rangle=0$, in order to achieve a singlet configuration on the $^1$O$_2$ subsystem for image 1. Following CASSCF, contour plots of the resulting orbitals are visualized in Fig.~\ref{fig:neb_avas_cas}b using NGL Viewer~\cite{ngl1, ngl2} with an isosurface value of 2.0, showing the characteristic $\pi$ and $\pi^*$ orbitals of O$_2$ which are involved in the reaction. The energies of subsequent images are calculated by initializing the molecular orbitals from the converged solution of the previous image. 

The resulting classically computed energies show intersecting potential energy surfaces typical of strongly correlated transition states, shown in Fig.~\ref{fig:neb_avas_cas}c. We note in particular the conical intersections observed around images 7 - 10, along with the local minimum at image 12, followed by a collapse to a closed shell singlet at image 16 to form dioxetane. Coefficients of the chemical statevector (configuration interaction (CI) coefficients) from converged CASSCF calculations for the (4, 6) AVAS subset lack a single large weight component for images 1 - 15, which indicates multireference character~\cite{bartlett12}. This is the case for all NEB images apart from the final (dioxetane), which is weakly correlated. Natural orbital populations ($\eta$)~\cite{szabo_ostlund} corresponding to eigenvalues of the 1-body reduced density matrix are shown in Fig.~\ref{fig:neb_avas_cas}a; $\eta$ values which deviate significantly from 2 (fully occupied) or 0 (empty) also indicate multireference character~\cite{bartlett12}. These indicators suggest strong electronic correlation; however, they are not sufficient to determine the degree of quantum entanglement contributing to this correlation, which is a focus of this work. We later use the term ``strong correlation'' to refer to the strong multireference character of the electronic wavefunction, but note that a comprehensive description of correlation structure requires more details than multireference indicators. 

For NEB images 1, 8, 12, and 16, the four largest weighted CI components (Slater determinants~\cite{szabo_ostlund}) from the respective CASSCF calculations are subsequently used to build quantum circuit ansätze, optimized in VQE, to represent these chemical states (discussed in the \hyperlink{qchem}{Quantum Computing Chemical States} subsection). The expectation values of ORDM elements can then be measured with respect to the state of these circuits. These experiments are presented in the \hyperlink{singlet}{$\langle S^2 \rangle = 0$ singlet} subsection of the \hyperlink{results}{Results} section.

In order to investigate characteristics of the chemical wavefunction that lead to orbital entanglement, we later consider alternative spin symmetries of the molecular statevector that can yield spin configurations not present in the singlet case. For this purpose, we relax the constraint of the $S^2$ total spin operator during the CASSCF optimizations. This results in $\langle S^2 \rangle = 2$ for NEB images 1 - 8 (the geometries are fixed to the NEB-DFT coordinates previously obtained), in which open shell spin configurations are present in the CI basis states (see Table~\ref{tab:sv_coeffs_gates_triplet}). Visualizations of the 4 energetically shallowest (O$_2$ $p$ projected) AVAS orbitals corresponding to an open shell triplet ($\langle S^2 \rangle = 2$) with equal numbers of spin up and spin down electrons ($s_z=0$) are shown above the correlation graphs of Fig.~\ref{fig:image1_image8_to2start}. In this case orbital energy levels are reordered (from SCF optimizations) relative to the singlet case, leading to a reordering of orbitals $i=0$ and $i=1$ for the open shell triplet (compare orbitals shown in the top panel of Fig.~\ref{fig:image1_image8_to2start} to Fig.~\ref{fig:neb_avas_cas}b). We note the CI coefficients obtained for this case also point to strong multireference character (see Table~\ref{tab:sv_coeffs_gates_triplet}) in the statevector due to lack of dominance of a single basis state, as observed for the singlet case. In this case, however, basis states with unpaired spin configurations are observed in the expansion, i.e. in this orbital basis the wavefunction exhibits correlations between open shell spin configurations. In the \hyperlink{triplet}{$\langle S^2 \rangle = 2$ triplet} subsection of the \hyperlink{results}{Results} section we show how these configurations lead to one-orbital quantum entanglement (in the sense described in the \hyperlink{methods}{Methods} subsection discussing orbital reduced density matrices with \hyperlink{ssrs}{fermionic SSRs}). 

\begin{table*}[t]
    \centering
    \adjustbox{max width=\textwidth}{
    \begin{tabular}{|c|c|c|c|c|}
        \hline
        NEB image & Quantum chemical statevector & PhasedX & $R_Z$ & ZZPhase \\[5pt]
        \hline
        1 & $\begin{array} {lcl} \ \ \ 0.6858464 \boldsymbol | \ (11)_0, (11)_1, (10)_2, (01)_3 \ \boldsymbol \rangle + 0.6858464 \boldsymbol | \ (11)_0, (11)_1, (01)_2, (10)_3 \ \boldsymbol \rangle \\+ \ 0.1720891 \boldsymbol | \ (10)_0, (01)_1, (11)_2, (11)_3 \ \boldsymbol \rangle + 0.1720891 \boldsymbol | \ (01)_0, (10)_1, (11)_2, (11)_3 \ \boldsymbol \rangle\end{array}$ & 22 & 11 & 4 \\[10pt]
        \hline
        8 & $\begin{array} {lcl} \ \ \ 0.7055596 \boldsymbol | \ (11)_0, (11)_1, (10)_2, (01)_3 \ \boldsymbol \rangle + 0.7055596 \boldsymbol | \ (11)_0, (11)_1, (01)_2, (10)_3 \ \boldsymbol \rangle \\- \ 0.0467516 \boldsymbol | \ (10)_0, (01)_1, (11)_2, (11)_3 \ \boldsymbol \rangle - 0.0467516 \boldsymbol | \ (01)_0, (10)_1, (11)_2, (11)_3 \ \boldsymbol \rangle\end{array}$ & 22 & 11 & 4
        \\[2pt]
        \hline
    \end{tabular}
    }
    \caption{Basis states of the VQE optimized statevector of 4 NEB images along the reaction path, in which the $S^2$ constraint on CASSCF optimizations were relaxed, producing an $\langle S^2 \rangle = 2$ (with total $s_z=0$) triplet. The orbital basis is arranged such that all basis states have the form $|\dots, (n^{\uparrow} \ n^{\downarrow})_i, (n^{\uparrow} \ n^{\downarrow})_{i+1}, \dots \rangle $, where $i$ labels the molecular orbitals (shown above the plots of Fig.~\ref{fig:image1_image8_to2start}), and $n^{\sigma}$ is the $\sigma$ spin orbital occupation which corresponds to a qubit state (with qubit index $2i$ ($2i+1$) for $\sigma=\ \uparrow \ (\downarrow)$) after JW transformation. Rightmost columns report the number of one-qubit PhasedX and $R_Z$ gates, and two-qubit ZZPhase gates (see Appendix C of the Supplementary Information for more details).}
    \label{tab:sv_coeffs_gates_triplet}
\end{table*}

\hypertarget{qchem}{\subsection*{Quantum Computing Chemical States}}\label{sec:methods:qchem}

\noindent Quantum calculations are performed using the \mbox{InQuanto} software package~\cite{inquanto_web, inquanto_docs} version 3.6, with circuits compiled using the architecture agnostic quantum software compiler \mbox{TKET}~\cite{tket20, tket_url}. We note that the JW transformation~\cite{jordan28} is used throughout. In terms of states, the JW transformation maps the 4 molecular orbital, 6 electron states (in a space of AVAS projected orbitals, as discussed in the \hyperlink{cchem}{Classical Computational Chemistry} subsection) to 8 qubit states with Hamming weight 6. To obtain quantum circuit representations of the chemical states, we use an ansatz to encode a subset of CI Slater determinants, followed by (classical) offline VQE optimization~\cite{vqe_opt} to refine the CI coefficients. For this VQE ansatz we consider the four largest weight Slater determinants observed in the CASSCF calculations. Linear combinations of these determinants (similar to selected CI \cite{szabo_ostlund}) are prepared using controlled Givens rotations, in a method developed as part of InQuanto and based on universality proofs of Givens rotations~\cite{anselmetti21, arrazola22}, implemented in \mbox{InQuanto}~\cite{inquanto_docs}. In this approach, the CI coefficients are directly related to gate angles, and VQE optimization leads to the correct (lowest energy) coefficients of basis vectors spanning the selected CI space. The VQE objective function is taken as the expectation value of the JW transformed second quantized chemical Hamiltonian \cite{mcardle20} consisting one- and two-body interactions indexed by spin orbitals. 

Following ideal VQE optimization with a classically evaluated objective function, total energies from VQE closely follow those obtained from CASSCF, as shown in Fig.~\ref{fig:neb_avas_cas}c. From the optimized VQE parameters, quantum circuit representations of the ground states of all NEB images of the reaction path are constructed for the (4, 6) AVAS subset. The resulting CI coefficients obtained from these circuits are presented in Table~\ref{tab:sv_coeffs_gates}. The CI coefficients match well those observed in CASSCF~\cite{ci_coeffs}. The quantum circuits are therefore representative of the chemical states for each NEB geometry along the reaction path. An example circuit is shown in Supplementary Fig. S1 (a), corresponding to image 1 of the NEB path in which the (rounded) gate angles are found by VQE such that the statevector of the circuit matches that for image 1 shown in Table~\ref{tab:sv_coeffs_gates}, and similarly (with different gate angles) for the other NEB images.

As mentioned in the \hyperlink{cchem}{Classical Computational Chemistry} subsection, CASSCF calculations are also performed with the $S^2$ constraint relaxed, resulting in the $s_z = 0$ component of the $\langle S^2 \rangle = 2$ triplet for NEB images representing the initial stages of the reaction. Following the same procedure as above, the largest weight Slater determinants from these CASSCF calculations are loaded into \mbox{InQuanto}. The gate angles are then optimized in VQE resulting in the chemical states reported in Table~\ref{tab:sv_coeffs_gates_triplet}. As can be seen, these states host open shell (molecular orbitals containing only 1 electron) configurations where the unfilled orbitals contain electrons with opposite spin (hence $s_z = 0$). For this case, NEB images 1 and 8 are selected to investigate orbital correlation and entanglement in the presence of such open shell configurations. An example circuit for the triplet is shown in Supplementary Fig. S1 (b), corresponding to image 1 of the NEB path, in which the VQE-optimized gate angles (rounded) resulted in the statevector for image 1 shown in Table~\ref{tab:sv_coeffs_gates_triplet}.

\hypertarget{corrent}{\subsection*{Measuring Correlation and Entanglement}} \label{sec:methods:qitchem}

\noindent This section describes our methods for quantifying total correlation and entanglement with a noisy quantum computer. First, we define our measures for total correlation and entanglement in the context of quantum chemistry, largely following previous work\cite{ding20,ding2020concept}. Then we relate those quantities to the measurements required from a quantum computer with and without the presence of fermionic superselection rules. Finally, we discuss a post-processing step for noise reduction.

Consider a quantum state described by a density matrix $\rho$ on a finite-dimensional Hilbert space $\mathcal{H}$. States form a convex set $D$ with the extremal points defining \emph{pure states} $\rho=\dyad{\psi}$ \cite{bengtsson17}. We denote the set of observables on $\mathcal{H}$ by $\mathcal{B}(\mathcal{H})$. Now consider a quantum system composed of two distinct subsystems $A$ and $B$ with Hilbert space $\mathcal{H} = \mathcal{H}_A \otimes \mathcal{H}_B$. The state of subsystem $A$ is described by the \emph{reduced density matrix} $\rho_{A} = \tr_B[\rho]$ (switch $A$ and $B$ for the state of $B$).  
The set of observables on the combined system has a corresponding tensor product structure $\mathcal{B}(\mathcal{H}_A) \otimes \mathcal{B}(\mathcal{H}_B)$. 
The expectation value of measuring a \mbox{local observable $\mathcal{O}(A) \in \mathcal{B}(\mathcal{H}_A)$ on subsystem $A$ is}
\vspace{.25em}
\begin{equation*}
    \langle \mathcal{O}(A) \rangle_{\rho_{A}} = \langle \mathcal{O}(A) \otimes \mathcal{I}(B) \rangle_{\rho} 
    = \tr[\rho (\mathcal{O}(A) \otimes \mathcal{I}(B))] ,
\end{equation*}

\noindent where $\mathcal{I}(B)$ is the identity operator acting on subsystem $B$ (and equivalent for local observables on $B$).

\emph{Uncorrelated states} are the states with density matrix $\rho$ satisfying \mbox{$\langle \mathcal{O}(A) \otimes \mathcal{O}(B) \rangle_\rho = \langle \mathcal{O}(A) \rangle_{\rho_A} \langle \mathcal{O}(B) \rangle_{\rho_B}$} for any local observables $\mathcal{O}(A)$ and $\mathcal{O}(B)$ on the respective subsystems. We denote the set of uncorrelated states as $D_0$. For distinguishable subsystems this is equivalent to writing the density matrix as a \emph{product state} $\rho = \rho_A \otimes \rho_B$.
\emph{Classically correlated states} are states which can be prepared as probabilistic ensembles of product states. They form the convex set of \emph{separable states} $D_\text{sep}$.

We quantify the \emph{total correlation} of a state $\rho$ with the quantum mutual information~\cite{rissler06, aliverti-piuri24, ding20, ding22, ding2020concept}
\begin{equation}\label{eq:mutual_information}
    I_{AB}(\rho) = \min_{\sigma\in D_0} S(\rho \Vert \sigma),
\end{equation}
where $S(\rho \Vert \sigma)$ is the quantum relative entropy. This can also be written as $I_{AB}(\rho) = S(\rho_A) + S(\rho_B) - S(\rho)$ with $S(\rho)=-\tr(\rho \log_2 \rho)$ the von Neumann entropy. The total correlation comprises a classical and a quantum contribution. Following \cite{ding20,ding2020concept}, \emph{entanglement} -- the quantum contribution to the total correlation -- is quantified by the minimal distance of $\rho$ to the set of separable states $D_\text{sep}$ as measured by the quantum relative entropy
\begin{equation}\label{eq:entanglement}
    E_{AB}(\rho) = \min_{\sigma \in D_\text{sep}} S(\rho \Vert \sigma).
\end{equation}
While the details of the minimization in those definitions are not relevant for this work, these definitions paint an intuitive geometric picture \cite{bengtsson17}. The total correlation is the minimal distance of $\rho$ to the set of uncorrelated states $D_0$ as measured by the quantum relative entropy. Entanglement is the minimal distance to the set of separable states. Note that both sets are defined with respect to a set of local observables on two subsystems.

The quantum information theoretic concepts discussed in the preceding paragraphs are developed for \emph{distinguishable} (sub)systems, and practical details surrounding the assumption of distinguishability are discussed in the next subsection. However in chemistry one considers correlations among orbitals hosting \emph{indistinguishable} electrons~\cite{ding20, aliverti-piuri24, balachandran13}. It has been argued \cite{ding20, ding2020concept} that care should be taken when quantifying total correlation and entanglement between orbitals to account for local fermionic symmetries, namely, fermionic superselection rules as discussed in the subsection titled ``\hyperlink{ssrs}{With fermionic superselection rules}''.

\vspace{0.5em}

\noindent \textbf{\textsf{Orbital reduced density matrices without fermionic \mbox{superselection} rules:}}\ We consider two types of subsystems $A$ of the total ground state molecular wavefunction: a single orbital $i$ and a pair of orbitals $(i, j)$ with $i, j = 1, 2, 3, 4$. In this section we only consider the global fermionic symmetries: fixed total number of electrons and spin symmetry. We denote by $\rho^{(1)}_i$  the one-orbital reduced density matrix (1-ORDM) for orbital $i$, where we trace out the remaining orbitals. We denote by $\rho^{(2)}_{i,j}$ the two-orbital reduced density matrix (2-ORDM) for orbital pair $(i, j)$, where we trace out the remaining orbital pairs. 

The Fock space of a single fermionic molecular orbital consists of 4 basis states corresponding to the 4 possible occupations of the two spin orbitals of the molecular orbital 
\begin{equation*}
    |\_ \ \_\rangle,|\_\downarrow\rangle,|\uparrow\_\rangle,|\uparrow\downarrow\rangle .
\end{equation*}
The ORDM elements are constructed from local fermionic operators acting on the molecular orbitals. First consider the 1-ORDM~\cite{boguslawski13, ding20, ding2020concept}

\begin{equation} \label{eqn:1ordm_mat}
    \begin{pmatrix} 
        \langle\mathcal{O}(i)_1\rangle & \langle\mathcal{O}(i)_2\rangle & \langle\mathcal{O}(i)_3\rangle & \langle\mathcal{O}(i)_4\rangle \\[3pt]
        \langle\mathcal{O}(i)_5\rangle & \langle\mathcal{O}(i)_6\rangle & \langle \mathcal{O}(i)_7\rangle & \langle\mathcal{O}(i)_8\rangle \\[3pt]
        \langle\mathcal{O}(i)_9\rangle & \langle\mathcal{O}(i)_{10}\rangle & \langle\mathcal{O}(i)_{11}\rangle & \langle\mathcal{O}(i)_{12}\rangle \\[3pt]
        \langle\mathcal{O}(i)_{13}\rangle & \langle\mathcal{O}(i)_{14}\rangle & \langle\mathcal{O}(i)_{15}\rangle & \langle\mathcal{O}(i)_{16}\rangle
    \end{pmatrix},
\end{equation}
where the 16 fermionic operators $\mathcal{O}(i)_{1\text{-}16}$ were reported in Boguslawski et al.~\cite{boguslawski13} and are listed in Appendix A of the Supplementary Information. The expectations are taken with respect to the total ground state molecular wavefunction in the given orbital basis. The $\mathcal{O}(i)_{1\text{-}16}$ are local operators applied to molecular orbital $i$ which correspond to each of the different mixings of the local orbital Fock states: the diagonal elements correspond to probabilities of orbital $i$ having the associated spin occupations, while off-diagonals represent correlations between the different occupations~\cite{boguslawski13, rissler06}. Spin and particle number are globally conserved for fermions, hence the physical 1-ORDM that preserves these quantities only has non-zero elements on its diagonal. For molecular orbital $i$ this becomes
 \begin{equation} \label{eqn:1ordm_enum}
    \rho^{(1)}_i = \begin{pmatrix} 
        \langle\mathcal{O}(i)_1\rangle & 0 & 0 & 0 \\ 

        0 & \langle\mathcal{O}(i)_6\rangle & 0 & 0 \\

        0 & 0 & \langle\mathcal{O}(i)_{11}\rangle & 0 \\

        0 & 0 & 0 & \langle\mathcal{O}(i)_{16}\rangle \\
    \end{pmatrix}.
\end{equation}
Given this bipartition inserting the molecular ground state into Eqs.~\eqref{eq:entanglement} and \eqref{eq:mutual_information} yields the one-orbital \mbox{entanglement} and correlation~\cite{ding2020concept,boguslawski13, boguslawski15, vedral97, rissler06}
\begin{align}
    E_i &= s1_i = -\sum_{\alpha} \omega_{\alpha, i} \log_2 \omega_{\alpha, i}, \label{eqn:s1}\\
    I_i &= 2 E_i,
\end{align}
where $\omega_{\alpha, i}$ is the $\alpha$-th eigenvalue of the $i$-th 1-ORDM, and we denote by $s1_i=S(\rho_i^{(1)})$ the one-orbital von Neumann entropy.

\begin{table*}[t]
    \centering
    \begin{tabular}{|c|c|}
        \hline
        $(n_e, s_z)$ & \\
        \hline
        (0, 0) & $^{1, 1}\langle \mathcal{O}(i)_1\mathcal{O}(j)_1 \rangle $ \\[2pt]
        \hline
        (1, -$\frac{1}{2}$) & 
        $\begin{matrix}
            ^{2, 2}\langle\mathcal{O}(i)_1\mathcal{O}(j)_6\rangle & \transparent{0.5}^{2, 3}\langle\mathcal{O}(i)_2\mathcal{O}(j)_5\rangle \\ 
            \transparent{0.5}^{3, 2}\langle\mathcal{O}(i)_5\mathcal{O}(j)_2\rangle & ^{3, 3}\langle\mathcal{O}(i)_6\mathcal{O}(j)_1\rangle \\[2pt]
        \end{matrix}$
        \\
        \hline
        (1, $\frac{1}{2}$) & 
        $\begin{matrix}
            ^{4, 4}\langle\mathcal{O}(i)_1\mathcal{O}(j)_{11}\rangle & \transparent{0.5}^{4, 5}\langle\mathcal{O}(i)_3\mathcal{O}(j)_9\rangle \\ 
            \transparent{0.5}^{5, 4}\langle\mathcal{O}(i)_9\mathcal{O}(j)_3\rangle & ^{5, 5}\langle\mathcal{O}(i)_{11}\mathcal{O}(j)_1\rangle \\[2pt]
        \end{matrix}$
        \\
        \hline
        (2, -1) & $^{6, 6}\langle \mathcal{O}(i)_6\mathcal{O}(j)_6 \rangle $ \\[2pt]
        \hline
        (2, 0) & 
        $\begin{matrix}
            ^{7, 7}\langle\mathcal{O}(i)_1\mathcal{O}(j)_{16}\rangle & \transparent{0.5}^{7, 8}\langle\mathcal{O}(i)_2\mathcal{O}(j)_{15}\rangle & \transparent{0.5}^{7, 9}\langle\mathcal{O}(i)_3\mathcal{O}(j)_{14}\rangle & \transparent{0.5}^{7, 10}\langle\mathcal{O}(i)_4\mathcal{O}(j)_{13}\rangle \\ 
            \transparent{0.5}^{8, 7}\langle\mathcal{O}(i)_5\mathcal{O}(j)_{12}\rangle & ^{8, 8}\langle\mathcal{O}(i)_6\mathcal{O}(j)_{11}\rangle & ^{8, 9}\langle\mathcal{O}(i)_7\mathcal{O}(j)_{10}\rangle & \transparent{0.5}^{8, 10}\langle\mathcal{O}(i)_8\mathcal{O}(j)_9\rangle \\
            \transparent{0.5}^{9, 7}\langle\mathcal{O}(i)_9\mathcal{O}(j)_8\rangle & ^{9, 8}\langle\mathcal{O}(i)_{10}\mathcal{O}(j)_7\rangle & ^{9, 9}\langle\mathcal{O}(i)_{11}\mathcal{O}(j)_6\rangle & \transparent{0.5}^{9, 10}\langle\mathcal{O}(i)_{12}\mathcal{O}(j)_5\rangle \\
            \transparent{0.5}^{10, 7}\langle\mathcal{O}(i)_{13}\mathcal{O}(j)_4\rangle & \transparent{0.5}^{10, 8}\langle\mathcal{O}(i)_{14}\mathcal{O}(j)_3\rangle & \transparent{0.5}^{10, 9}\langle\mathcal{O}(i)_{15}\mathcal{O}(j)_2\rangle & ^{10, 10}\langle\mathcal{O}(i)_{16}\mathcal{O}(j)_1\rangle \\[2pt]
        \end{matrix}$
        \\
        \hline
        (2, 1) & $^{11, 11}\langle \mathcal{O}(i)_{11}\mathcal{O}(j)_{11} \rangle $ \\[2pt]
        \hline
        (3, -$\frac{1}{2}$) & 
        $\begin{matrix}
            ^{12, 12}\langle\mathcal{O}(i)_6\mathcal{O}(j)_{16}\rangle & \transparent{0.5}^{12, 13}\langle\mathcal{O}(i)_8\mathcal{O}(j)_{14}\rangle \\ 
            \transparent{0.5}^{13, 12}\langle\mathcal{O}(i)_{14}\mathcal{O}(j)_8\rangle & ^{13, 13}\langle\mathcal{O}(i)_{16}\mathcal{O}(j)_6\rangle \\[2pt]
        \end{matrix}$
        \\
        \hline
        (3, $\frac{1}{2}$) & 
        $\begin{matrix}
            ^{14, 14}\langle\mathcal{O}(i)_{11}\mathcal{O}(j)_{16}\rangle & \transparent{0.5}^{14, 15}\langle\mathcal{O}(i)_{12}\mathcal{O}(j)_{15}\rangle \\ 
            \transparent{0.5}^{15, 14}\langle\mathcal{O}(i)_{15}\mathcal{O}(j)_{12}\rangle & ^{15, 15}\langle\mathcal{O}(i)_{16}\mathcal{O}(j)_{11}\rangle \\[2pt]
        \end{matrix}$
        \\
        \hline
        (4, 0) & $^{16, 16}\langle \mathcal{O}(i)_{16}\mathcal{O}(j)_{16} \rangle $ \\[2pt]
        \hline
    \end{tabular}
    \caption{Non-zero elements of the 2-ORDM $\rho^{(2)}_{i, j}$ for molecular orbitals $i, j$. All other elements are null by global fermionic symmetries. Each operator $\mathcal{O}$ corresponds to one of the 16 terms labelled in Eq.~\eqref{eqn:1ordm_mat}, denoted by its subscript. The superscripts outside the angular brackets denote the position of the element in the 16$\times$16 matrix. Faded terms correspond to those elements which are set to 0 when taking into account local fermionic SSRs. Left column labels the local electron number ($n_e$) and spin ($s_z$) sectors.}
    \label{tab:2ordm}
\end{table*}

For the 16$\times$16 2-ORDM, global fermionic symmetries result in 36 non-zero elements~\cite{boguslawski13}, labelled by the product of members of the pool of 16 operators $\mathcal{O}(i)_{1\text{-}16}$. The 2-ORDM matrix elements can be grouped into sectors corresponding to local quantum numbers of the two-orbital pair subsystem, and the total matrix has been shown in previous works~\cite{boguslawski13, ding2020concept}. For completeness, in Table~\ref{tab:2ordm} we show the matrix elements associated with non-zero sectors. With this bipartition the total two-orbital correlation is~\cite{rissler06}
\begin{equation} \label{eqn:mi}
    I_{i,j} = \frac{1}{2}(s1_i + s1_j - s2_{i,j})(1 - \delta_{i,j})
\end{equation}
with
\begin{equation}\label{eqn:s2}
    s2_{i,j} = S\left(\rho_{i,j}^{(2)}\right) = -\sum_{\alpha} \omega_{\alpha, i,j} \log_2 \omega_{\alpha, i,j}
\end{equation}
the two-orbital von Neumann entropy and $\omega_{\alpha, i, j}$ the $\alpha$-th eigenvalue of the 2-ORDM for orbital pair $(i, j)$. Equation~\eqref{eq:entanglement} for two-orbital entanglement becomes a high-dimensional and highly non-trivial optimization problem~\cite{ding2020concept, ding22}. Here we do not consider two-orbital entanglement.

The ORDM elements are constructed from the fermionic operators $\mathcal{O}(i)_{1\text{-}16}$ (defined in Appendix A of the Supplementary Information) and their products, which are encoded using the JW transformation. For more details on the resulting Pauli strings, see Appendix B of the Supplementary Information. We measure the entries of the 1-ORDM and 2-ORDM with respect to the VQE-approximated ground states prepared on the quantum computer. In Boguslawski et al.~\cite{boguslawski13} it was observed that the 36 elements of the 2-ORDM can be reduced to 26 classical expectation evaluations due to symmetry of the matrix. This would naively translate to 156 evaluations for the 6 orbitals pairs generated from the 4 AVAS orbitals. The Pauli strings defining the ORDM elements can also be grouped into commuting sets to reduce the total number of measurements~\cite{cowtan20, jena19, tket_url, miller24}. Here, searches for commuting sets are performed within a set of Pauli operators corresponding to a given ORDM.
Interestingly, we find that the 36 expectation values per 2-ORDM can be reduced to approximately 6 measurable circuits per orbital pair, defining circuits for their expectation value measurements (with respect to the previously found ground state circuits) and a partitioning of measurable expectation values according to which Pauli operator strings commute (and can thus be measured simultaneously). This translates to 35 circuits in total to measure all 2-ORDMs. For the 1-ORDM, whose diagonals are essentially number operator products and thus strings of Pauli Z rotations after JW transformation, the number of measured circuits reduces to 1 per 1-ORDM after grouping into commuting sets. This results in 39 measurement circuits altogether to extract all 1-ORDMs and 2-ORDMs. They can be reduced even further when superselection rules are taken into account, discussed in the next subsection. 
\newline

\noindent \hypertarget{ssrs}{\textbf{\textsf{With fermionic superselection rules:}}}\ Recently, it has been argued that inferring correlations from the spectra of the ORDM can overestimate the correlation due to lack of accounting for SSRs~\cite{ding20, ding2020concept, ding22, balachandran13}. For fermions, SSRs forbid symmetry-breaking superpositions of basis states, which amounts to restricting the local algebra of observables to operators that preserve the relevant symmetries. While global fermionic symmetries are already accounted for in the ORDMs discussed in the preceding subsection, SSRs additionally require conservation of \textit{local} fermionic symmetries. The SSRs pertinent to this work correspond to fermion parity and number~\cite{wick52, wick70}. Breaking of parity SSR corresponds to superpositions of pure states with even and odd local parity leading to violation of the no-signaling theorem~\cite{friis16}. The number SSR corresponds to superpositions of pure states with different numbers of fermions (regardless of parity) in the local subsystems. Breaking the number SSR is not fundamentally forbidden but it is reasonable to assume that it holds in typical quantum chemistry settings~\cite{ding2020concept}. We only consider number SSR because it accounts for parity SSR in the total correlation and entanglement relevant to this work~\cite{ding2020concept}.

SSRs restrict the admissible local observables on subsystems $A$ and $B$. Namely, physical local observables $\mathcal{O}_{\mathcal{F}}(A), \mathcal{O}_{\mathcal{F}}(B)$ are block-diagonal in the eigenbasis of the quantity conserved by the SSR. They belong to a restricted algebra of local fermionic operators $\mathcal{F}_{A}, \mathcal{F}_{B}$, respectively, which satisfy this local symmetry. Starting from a general observable not necessarily respecting the SSR one recovers a physical local observable by projecting onto the eigenspaces of the conserved quantity. Clearly, the set of local observables respecting the SSR is a subset of all observables. The restriction carries over to the set of accessible states.
A physical state $\rho^{\text{SSR}}$ obeying fermionic SSRs is block-diagonal in the eigenbasis of the observable that is locally conserved under the SSR. If one starts from an unphysical state $\rho$ not respecting the SSR, the physical part of that state can be recovered by projecting onto the eigensubspace of the SSR-conserved property~\cite{ding2020concept}. 

Recall the definition of uncorrelated states in terms of a factorization of the tensor product structure of the local observables. Since the admissible local observables under the SSR are a subset of all local observables, we have that more states can satisfy the factorization condition defining an uncorrelated state. The net effect of SSR inclusion is to \emph{increase} the set of uncorrelated states:
\begin{equation*}
    \begin{split}
        D_0^\text{SSR} &= \Big\{\rho : 
        \langle \mathcal{O}_{\mathcal{F}}(A) \otimes \mathcal{O}_{\mathcal{F}}(B) \rangle_{\rho} \\
        &\quad = \langle \mathcal{O}_{\mathcal{F}}(A) \rangle_{\rho_A} \langle \mathcal{O}_{\mathcal{F}}(B) \rangle_{\rho_B}, 
        \\
        &\quad \forall \mathcal{O}_{\mathcal{F}}(A) \in \mathcal{F}_A, \mathcal{O}_{\mathcal{F}}(B) \in \mathcal{F}_B \Big\} \supseteq D_0.
    \end{split}
\end{equation*}
This also introduces a larger set of separable states $D_\text{sep}^\text{SSR} \supseteq D_\text{sep}$, defined as the convex hull of $D_0^\text{SSR}$. To compute SSR-corrected total correlation and entanglement for a state $\rho$ one inserts the physical state $\rho^\text{SSR}$ obtained from projection into Eqs.~\eqref{eq:mutual_information}, \eqref{eq:entanglement}~\cite{ding2020concept}.
Considering again the geometric picture of quantum states and correlation measures as distance metrics on this geometry, this shows that neglecting fermionic SSRs can indeed lead to overestimation of fermionic correlations: without SSRs the minimal distance between a state in question and the subset of uncorrelated states $D_0$ is in general too large, because that subset is too small. 

In the context of this work, inclusion of SSRs corresponds to setting certain ORDM elements to zero~\cite{ding2020concept}. While the diagonal elements in $\rho^{(1)}_i$ do not violate local SSRs by definition, the $\rho^{(2)}_{i, j}$ matrix is modified such that 18 additional elements are fixed to 0 resulting in a SSR-modified 2-ORDM $\rho^{(2),SSR}_{i, j}$. These elements are shown in a faded font in Table~\ref{tab:2ordm}. 

Closed-form expressions for the one-orbital total correlation and entanglement of orbital $i$ can be derived by considering the factorization of the Fock space and the resulting form of the total physical state $\rho^{\text{SSR}}$, whose eigenvalues become $\omega_{1,i}$, $\omega_{2,i} + \omega_{3,i}$, and $\omega_{4,i}$~\cite{ding2020concept}. Recall the expression for mutual information Eq.~\eqref{eq:mutual_information}; for one-orbital correlation the $A$ and $B$ subsystems refer to $i$ and the complement set of orbitals $/\{i\}$, respectively. Then, noting that the reduced density matrices for those subsystems share the same eigenvalues as $\rho_i^{(1)}$ due to symmetry~\cite{ding2020concept}, the total correlation (generally expressed as $S(\rho_A) + S(\rho_B) - S(\rho)$) reads in its SSR-corrected form for one orbital as

\begin{equation} \label{eqn:ssr_I}
    \begin{split}
        I^{\text{SSR}}_i = & \ \omega_{1,i}\log_2\omega_{1,i} 
        \\&+ (\omega_{2,i} + \omega_{3,i})\log_2(\omega_{2,i} +\omega_{3,i}) 
        \\&+ \omega_{4,i}\log_2\omega_{4,i} 
        \\&- 2\sum_{\alpha=1}^4 \omega_{\alpha,i}\log_2\omega_{\alpha,i},
    \end{split}
\end{equation}

\noindent while for the SSR-corrected one-orbital entanglement, the only contributions that survive correspond to single occupations \cite{ding22}. The one-orbital von Neumann entropy then only has contributions from $\omega_{2,i}$ and $\omega_{3,i}$, leading to

\begin{equation} \label{eqn:ssr_E}
    \begin{split}
        E^{\text{SSR}}_i = & \ (\omega_{2,i} + \omega_{3,i})\log_2(\omega_{2,i} +\omega_{3,i}) 
        \\&- \omega_{2,i}\log_2\omega_{2,i} 
        \\&- \omega_{3,i}\log_2\omega_{3,i}.
    \end{split}
\end{equation}
Note that the spectrum of the \mbox{1-ORDM} $\rho^{(1)}_i$ (indexed by $\alpha$) remains sufficient~\cite{ding2020concept, amosov16}. Since $\rho^{(1)}_i$ in Eq.~\eqref{eqn:1ordm_enum} is diagonal, the terms in Eqs.~\eqref{eqn:ssr_I} and~\eqref{eqn:ssr_E} corresponding to eigenvalues $\alpha = 1, 2, 3, 4$ could be replaced respectively with expectation values of operators $\mathcal{O}(i)_1, \mathcal{O}(i)_6, \mathcal{O}(i)_{11}, \mathcal{O}(i)_{16}$. For the SSR-corrected two-orbital total correlation we use Eq.~\eqref{eqn:mi} after replacing the eigenvalues $\omega_{\alpha,i,j}$ with the eigenvalues of the SSR-corrected 2-ORDM $\rho^{(2),SSR}_{i, j}$.

In terms of the number of circuits required, accounting for SSRs leads to only 3 circuits per 2-ORDM after grouping into commuting sets of Pauli operators. This results from the reduced number of non-zero matrix elements (non-faded terms in Table~\ref{tab:2ordm}). By inspection of the JW-transformed operators in Appendix B of the Supplementary Information, it can be seen that after accounting for SSRs the remaining operator products to build the 2-ORDM involve commutable strings of Pauli X and Y rotations, as well as Z strings. Taking orbital pair ($i=0$, $j=1$) as an example (involving qubits $q=0, \dots, 3$), the 3 commuting sets of Pauli strings after accounting for SSRs are
\begin{equation*}
\begin{split}
    \boldsymbol\{ &Z_0, Z_1, Z_2, Z_3, Z_0Z_1, Z_0Z_2, Z_0Z_3, Z_1Z_2, Z_1Z_3, Z_2Z_3, 
    \\&Z_0Z_1Z_2, Z_0Z_1Z_3, Z_0Z_2Z_3, Z_1Z_2Z_3, Z_0Z_1Z_2Z_3 \boldsymbol\} ,\\
    \\\boldsymbol\{&Y_0Y_1Y_2Y_3, X_0X_1X_2X_3, X_0Y_1X_2Y_3, X_0Y_1Y_2X_3, 
    \\&X_0X_1Y_2Y_3, Y_0X_1X_2Y_3, Y_0Y_1X_2X_3, Y_0X_1Y_2X_3 \boldsymbol\} ,\\
    \\\boldsymbol\{&X_0Y_1Y_2Y_3, X_0Y_1X_2X_3, X_0X_1X_2Y_3, Y_0X_1Y_2Y_3, 
    \\&Y_0X_1X_2X_3, Y_0Y_1X_2Y_3, Y_0Y_1Y_2X_3, X_0X_1Y_2X_3 \boldsymbol\} ,
\end{split}
\end{equation*}
where $P_q$ is applied to qubit $q$ and $P \in \{X, Y, Z\}$. Each set can be measured with 1 circuit. Combining with 1 measurement circuit for each 1-ORDM, this results in 22 measurement circuits altogether for all 1-ORDMs and 2-ORDMs when accounting for SSRs. We note that the number of 1-ORDMs scales linearly with system size (or number of qubits $n$), translating to $O(n/2)$ measurable elements (ignoring grouping into commuting sets) to obtain all one-orbital entropies. The number of 2-ORDMs equals the number of unique orbital pairs $(n_{\text{mo}}^2 - n_{\text{mo}}^{})/2$, where $n_{\text{mo}} = n/2$ is the number of molecular orbitals); as each 2-ORDM is of constant size, the number of 2-ORDM elements to measure all two-orbital entropies scales as $O(n^2/8- n/4)$.

We also comment on the use of JW transformation to represent the ORDM elements. In this encoding, products of Pauli Z operators account for fermionic exchange (see Appendix B of the Supplementary Information); for the 2-ORDM these Pauli Z products are non-local (spanning qubits outside those specified by the molecular orbital indexes) for non-adjacent orbital pairs ($\abs{i - j} > 1$). Different schemes which encode properties other than fermionic occupation (e.g. parity encoding) can lead to different noise profiles for the measurable circuits due to absence of these Pauli Z strings. However, whether alternative encoding schemes lead to overall improvement in accuracy depends on the resulting circuit resources (e.g. depth) and hence requires further investigation, which we leave to future work. 
\newline

\noindent \textbf{\textsf{Noise reduction:}}\ In computing entropies we expect our computations to be very sensitive to noise in the measured 1-ORDM and 2-ORDM matrices. 
This noise can occur both as statistical noise from measurement and noise from quantum errors in the hardware.
We apply two steps of post-processing to mitigate the effects of noise. The first is a filtering step that removes the small singular values of the measured matrices below a `hard-threshold'~\cite{donoho2013optimal}. The second is a correction step that finds the closest physically valid density matrix to the ORDM matrix we have obtained~\cite{smolin2012efficient}.

Statistical noise, resulting from using a finite number of shots in the measurement of the ORDM matrix elements, can be handled by hard-thresholding the small singular values of the noisy matrix. 
We assume that when trying to measure a $d \times d$ matrix $M$, we actually measure $\widetilde{M} = M + \sigma N$ where the elements of $N$ are independent, identically distributed, zero-mean random variables.
When the noise level, $\sigma$, is known, then  $({4} / {\sqrt{3}}) \, \sqrt{d} \, \sigma$ is the optimal threshold for removing small, noise induced singular values ~\cite{donoho2013optimal}. 
Here we set $\sigma = 1/\sqrt{N_\textrm{shots}}$ where $N_\textrm{shots}$ is the number of measurement shots.
An unusual aspect of our noisy matrices is that we do not measure all the entries of the matrix since for symmetry reasons many of them must be zero. We find that it is then appropriate to reduce the hard-threshold to 
\begin{equation}
    ({4} / {\sqrt{3}}) \, \sqrt{d R} \, \sigma ,
\end{equation}
where $R$ is the density of non-zero entries (see Fig. S2 and Appendix D of the Supplementary Information for more details). The number of non-zero entries $R \, d^2$ is 4 for the 1-ORDMs and 18 (36) for the 2-ORDMs with (without) SSRs.
We compute the singular value decomposition of the measured matrices, which has computational cost $O(d^3)$ for $d \times d$ matrices, and we set any singular values smaller than our threshold to zero.

The hard-thresholding of the singular values can also be a useful heuristic to help mitigate circuit errors. In general, quantum errors can act in a much more complicated way. However, in some situations a good approximation is achieved by assuming a global depolarising model~\cite{cai2023quantum, qin2023error} where the noisy state $\rho$ is related to the ideal state $\ket{\psi}$ by
\begin{equation}
    \rho \approx (1 - p) \ketbra{\psi}{\psi} + p \frac{\mathbb{I}^{\otimes n}}{2^n},
\end{equation}
with $\mathbb{I}$ the single qubit identity matrix and where $p$ is the depolarising parameter of the effective global depolarising channel. In this case, the reduced density matrices from a partial trace of $\rho$ will also have the form
\begin{equation}
    \tr_B \rho \approx (1 - p) \, \tr_B \ketbra{\psi}{\psi} + p \frac{\mathbb{I}^{\otimes m}}{2^m},
\end{equation}
where $m = n - |B|$ is the number of remaining qubits, with $|B|$ the number of qubits traced out. If $p$ is sufficiently small and the hard-threshold is set above $p$, then the circuit noise contributions to $\tr_B \rho$ will be removed.

The second step we apply is a correction step. The ORDM matrices we measure are not guaranteed to be physically valid density matrices. To find the closest physical state we use a fast, maximum likelihood based method that assumes Gaussian noise and has computational cost $O(d^3)$ for $d \times d$ matrices~\cite{smolin2012efficient}.

\hypertarget{results}{\section*{Results}}

\begin{figure*}[t]
    \centering
    \includegraphics[width=0.98\textwidth]{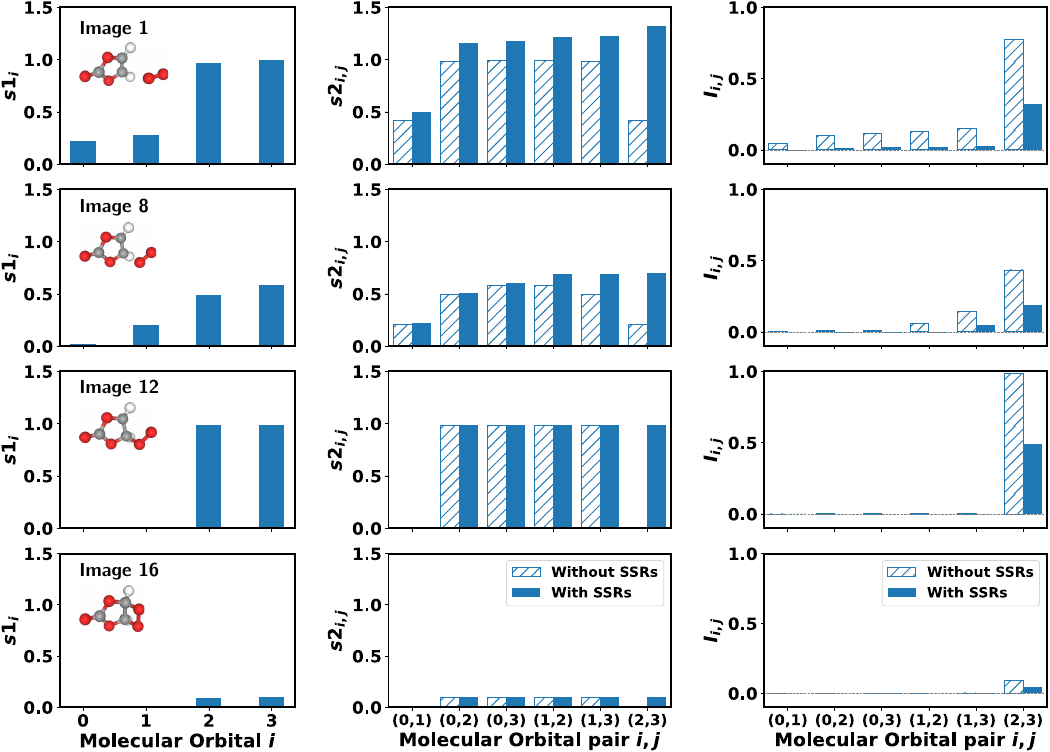}
    \caption{One-orbital von Neumann entropies, two-orbital von Neumann entropies, and two-orbital mutual information, for 4 images of the NEB reaction path in which the CASSCF optimization is constrained to the $\langle S^2 \rangle = 0$ singlet. Obtained using ideal noiseless simulations. For each row of panels, the corresponding molecular structure is displayed within the $s1_i$ plot. When accounting for SSRs, the one-orbital entanglement $E^{\text{SSR}}_i$ (not shown) is 0, and in this limit $I^{\text{SSR}}_i = s1_i$~\cite{ding2020concept}. Orbital indexes correspond to orbitals labelled in Fig.~\ref{fig:neb_avas_cas}b. When accounting for SSRs, two-orbital mutual information on vertical axis of rightmost panels corresponds to $I^{\text{SSR}}_{i,j}$ (otherwise $I_{i,j}$).
    }
    \label{fig:4images_mi_1o2start}
\end{figure*}

\noindent In this section, we present our analysis of the orbital correlations in the VC + $^1$O$_2$ $\rightarrow$ dioxetane reaction. 
This includes ideal (noiseless) computations, which give us a clear picture of the behavior we expect to see, as well as yielding interesting insights into the chemistry. 
Additionally, we demonstrate that we are able to effectively reproduce these estimates of the correlations from experiments on quantum hardware.
Our results are split between the singlet and triplet molecular states (with $\langle S^2 \rangle$ values for the total spin operator in the subsection titles). 

\subsection*{\texorpdfstring{$\langle S^2 \rangle = 0$ singlet}{Singlet}}\hypertarget{singlet}{\label{subsec:singlet}}

\noindent Here we study the singlet spin configuration.
We first consider ideal (noiseless) computations of the ORDMs, obtained by calculating expectation values using statevector simulations of the same circuits run on hardware.
Following that, we show our results from experiments on quantum hardware.
\newline

\noindent \textbf{\textsf{Ideal results:}}\ In Fig.~\ref{fig:4images_mi_1o2start}, we observe the dominance of orbitals $i=2$ and $i=3$ in terms of total one-orbital correlations throughout the reaction path. Taking either of these orbitals as subsystems in a quantum information theoretic sense, these results indicate a high degree of correlation between the respective orbital subsystem and the remaining orbitals. The two-orbital von Neumann entropies $s2_{i, j}$ and resulting mutual information values reflect this picture: the largest $s2_{i, j}$ values occur when either $i$ or $j$ are 2 or 3, and the orbital pair $i=2, j=3$ hosts the largest mutual information. 

We also note the impact of accounting for fermionic SSRs. In this case, all one-orbital entanglements $E^{\text{SSR}}_i$ defined by Eq.~\eqref{eqn:ssr_E} are 0 throughout the reaction (not shown in Fig.~\ref{fig:4images_mi_1o2start}, for clarity). In this case the total one-orbital correlation when accounting for SSRs is simply related to the non-SSR value as $I^{\text{SSR}}_i = \frac{I_i}{2}$~\cite{ding2020concept}. By inspecting the terms of Eq.~\eqref{eqn:ssr_E}, we see that contributions to the von Neumann entropy from 1-ORDM diagonals $\alpha=2, 3$ are 0. The corresponding diagonal values refer to probabilities for orbital $i$ to have single spin occupations (specifically, those diagonals refer to operators $\mathcal{O}(i)_6 = \hat{n}_{i, \downarrow} - \hat{n}_{i, \uparrow}\hat{n}_{i, \downarrow}$ and $\mathcal{O}(i)_{11} = \hat{n}_{i, \uparrow} - \hat{n}_{i, \uparrow}\hat{n}_{i, \downarrow}$, where $\hat{n}_{i, \sigma}$ is the number operator for spin orbital $\sigma$ of molecular orbital $i$). By noting those elements evaluate (ideally) to 0 for this case, and comparing to the statevector expansions shown in Table~\ref{tab:sv_coeffs_gates}, it is clear that vanishing orbital entanglement (after accounting for SSRs) is the result of only closed shell (spin paired) configurations in the basis states: 0 probabilities of single spin occupations in an orbital ultimately lead to Eq.~\eqref{eqn:ssr_E} evaluating to 0. As previously shown~\cite{ding2020concept}, for 0 entanglement
the total orbital correlation $I^{\text{SSR}}_i$ accounting for SSRs equals the one-orbital von Neumann entropy (Eq.~\eqref{eqn:s1}), which again is clear by inspection of Eq.~\eqref{eqn:ssr_I}. 

Regarding the impact of SSRs on the orbital pairs $(i, j)$, we note the two-orbital von Neumann entropies $s2_{i,j}$ are generally lower without accounting for SSRs for all orbitals pairs in NEB images 1 and 8, whereas they are identical for all pairs apart from (2, 3) in NEB images 12 and 16. Considering mutual information, noiseless values of $I_{i,j}$ are always higher than $I^{\text{SSR}}_{i,j}$, as expected for non-maximally entangled orbital pairs~\cite{ding2020concept}. This highlights the overestimation of correlation when not accounting for SSRs.
\newline
\newline

\begin{figure*}[t]
    \centering
    \includegraphics[width=\textwidth]{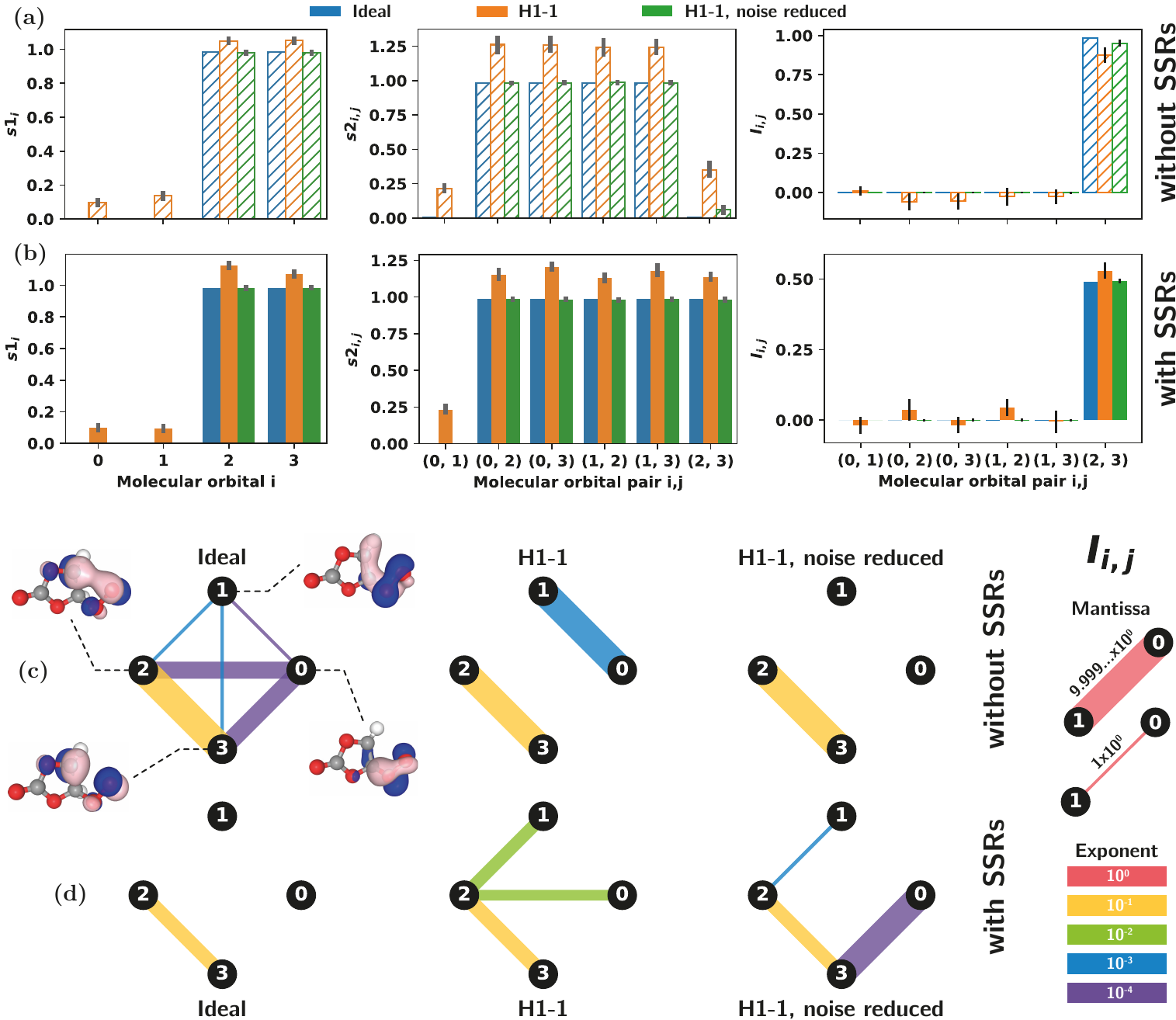}
    \caption{
    Orbital entropies and mutual informations for image 12 with and without superselection rules, obtained from experiments on the Quantinuum H1 trapped-ion quantum computer. 
    Single orbital entropies, $s1_i$, two orbital entropies, $s2_{i,j}$, and mutual information, $I_{i,j}$ are plotted for each molecular orbital or pair of orbitals without superselection rules in (a) and with superselection rules in (b). In each plot ideal values obtained from statevector simulations are plotted along with the experimentally measured values, both with and without noise reduction. Without noise reduction many of the estimated mutual informations have unphysical negative values.
    Error bars are obtained by bootstrap resampling the measurement shots 1000 times.
    Mutual information values are visualized as edges on a graph where the nodes are the four molecular orbitals. These are shown without superselection rules in (c) and with superselection rules in (d).
    Values are shown over many orders of magnitude by indicating the exponent through the color of the edge and the mantissa with its thickness.
    In (c) and (d) erroneous negative values are excluded and only positive values are displayed.}
    \label{fig:hardware-results-singlet}
\end{figure*}

\noindent \textbf{\textsf{Hardware results:}}\ To assess how well we are able to reconstruct the entropies and mutual information on quantum hardware we focus on image 12 with and without SSRs. The 1-ORDM and 2-ORDM are obtained from measurements on the Quantinuum H1-1 trapped-ion quantum computer~\cite{h11}. Each mutually commuting set of Pauli strings is measured using 10,000 shots~\cite{em_calcs}.

In Fig.~\ref{fig:hardware-results-singlet} we plot the single orbital entropies, $s1_i$, two orbital entropies, $s2_{i,j}$, and mutual information, $I_{i,j}$ with and without superselection rules. Raw data from the device is plotted alongside post-processed data using the noise reduction strategy described in the \hyperlink{methods}{Methods} section. To compute entropies from the raw data we simply diagonalize the measured matrices, remove any negative eigenvalues and rescale the spectrum to sum to one.

We see that the noise reduction strategy is highly effective. In the raw device data entropies are consistently overestimated, as we would expect for states with lower purity due to noise. In most cases this additional entropy is very well removed by thresholding the small singular values of the measured matrices.
Comparing the mutual information values we obtain, in Fig.~\ref{fig:hardware-results-singlet}(c) and (d) we see that the noise-reduction performs very well at removing relatively large spurious values in the raw data. However, small errors can still remain as is clear in (d) and the noise reduced data can miss fine details in the mutual information pattern as is evidenced in (c).

\subsection*{\texorpdfstring{$\langle S^2 \rangle = 2$ triplet}{Triplet}}\hypertarget{triplet}{\label{subsec:triplet}}

\begin{figure*}[t]
    \centering
    \includegraphics[width=\textwidth]{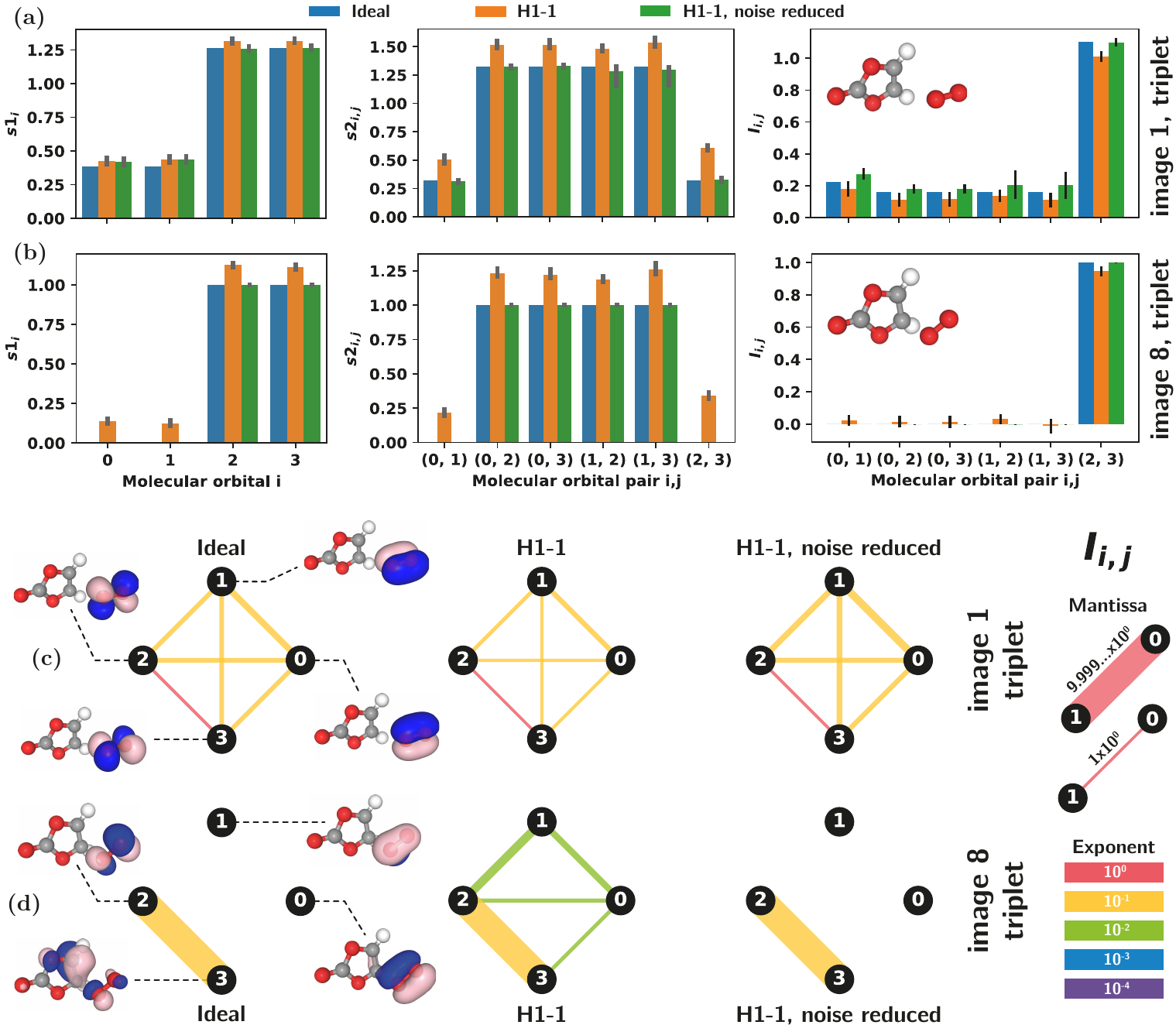}
    \caption{
    Orbital entropies and mutual informations obtained from the Quantinuum H1 trapped-ion quantum computer for image 1 and image 8 of the NEB reaction path in which the CASSCF wavefunction corresponds to the $\langle S^2 \rangle = 2$ triplet. Superselection rules are applied to both images.
    Single orbital entropies, $s1_i$, two orbital entropies, $s2_{i,j}$, and mutual information, $I_{i,j}$ are plotted for each molecular orbital or pair of orbitals for (a) image 1 and (b) image 8. In each plot ideal values (obtained from statevector simulations) are plotted along with the experimentally measured values, both with and without noise reduction. 
    Error bars are obtained by bootstrap resampling the measurement shots 1000 times.
    Mutual information values are visualized as edges on a graph where the nodes are the four molecular orbitals for (c) image 1 and (d) image 8.
    Values are shown over many orders of magnitude by indicating the exponent through the color of the edge and the mantissa with its thickness.
    In (c) and (d) any negative values are excluded and only positive values are displayed.}
    \label{fig:hardware-results-triplet}
\end{figure*}

\begin{figure*}[ht]
    \centering
    \includegraphics[width=0.85\textwidth]{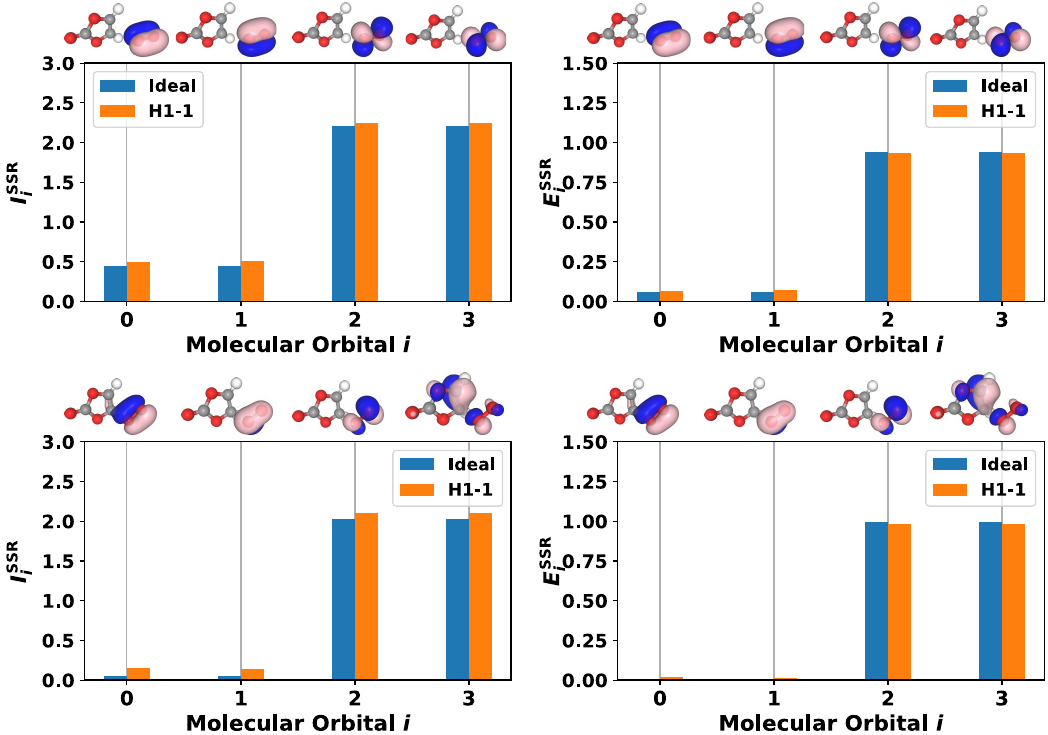}
    \caption{Total one-orbital correlation (left) and one-orbital entanglement (right) accounting for SSRs (using Eqs.~\eqref{eqn:ssr_I} and~\eqref{eqn:ssr_E}), for image 1 (top) and image 8 (bottom) of the NEB path, for the $s_z = 0$ triplet ($\langle S^2 \rangle = 2$). Orbitals representing the indexes of the horizontal axes are visualized for both NEB images and shown above each graph. Orange bars represent measurement data from the Quantinuum H1-1 trapped-ion quantum computer.}
    \label{fig:image1_image8_to2start}
\end{figure*}

\noindent To highlight the interesting role open shell configurations play in the orbital entanglement we also study the triplet spin configurations for images 1 and 8. Here all the data presented has SSRs enforced and, again, hardware data is collected from H1-1~\cite{h11} using 10,000 measurement shots per commuting Pauli set.

Fig.~\ref{fig:hardware-results-triplet} plots the single orbital entropies, $s1_i$, two orbital entropies, $s2_{i,j}$, and mutual information, $I_{i,j}$ for image 1 and image 8 in the triplet spin configuration, showing the ideal values along with the raw hardware data and the noise reduced hardware results.
Again we see the noise reduction strategy is effective at removing excess entropy in the raw hardware data.
In the case of the image 1 triplet data we see relatively large (and sometimes asymmetrical) error bars in the noise reduced $s2_{i,j}$ and $I_{i,j}$ estimates. This seems to arise from some of the measured singular values of the 2-ORDMs being close to the hard-threshold we use, so that some of the bootstrap resamples incorrectly drop those singular values.

Different from the singlet case (see the \hyperlink{singlet}{$\langle S^2 \rangle = 0$ singlet} subsection), where the total one-orbital correlation contained no contribution from entanglement, when accounting for SSRs, we see that entanglement contributes significantly to the one-orbital correlation for the triplet, Fig.~\ref{fig:image1_image8_to2start}. In the singlet case all basis states (see Table~\ref{tab:sv_coeffs_gates}) correspond to closed shell occupation configurations (all molecular orbitals are either empty or filled with two electrons). Whereas for the triplet, the wavefunction contains basis states which have spin-unpaired (open shell) molecular orbitals, and non-zero one-orbital entanglement is obtained. This effect can be easily understood by inspecting Eq.~\eqref{eqn:ssr_E}, which shows that contributions from 1-ORDM eigenvalues $\omega_{2,i}$ and $\omega_{3,i}$ must be significant for $E^{\text{SSR}}_i$ to be non-negligible. Note that the second and third diagonals of the 1-ORDM in Eq.~\eqref{eqn:1ordm_enum} represent probabilities for single electron occupation. We also mention that the $s_z = \pm 1$ triplet component (not shown here) would yield either $\omega_{2,i}=0$ or $\omega_{3,i}=0$ (but not both) in Eq.~\eqref{eqn:ssr_E}, resulting again in $E^{\text{SSR}}_i = 0$. This is consistent with the formalism of Ding et al.~\cite{ding2020concept}, and our results further emphasize that \emph{both} $\omega_{2,i}$ and $\omega_{3,i}$ eigenvalues should be non-zero in order for one-orbital entanglement to be non-negligible, which is accomplished in this case by open shell orbitals with opposite spin yielding the $s_z = 0$ component of the triplet.

\section*{Discussion}\label{sec:conclusions}

\noindent In summary, we have shown that orbital correlation and entanglement can be accurately reconstructed from measurements on a current trapped-ion quantum computer. 
As a model system we considered the VC + $^1$O$_2$ $\rightarrow$ dioxetane reaction. Molecular orbitals were constructed from a pipeline involving first the determination of atomic geometries using the NEB method, followed by an AVAS projection of the $p$ orbitals of the O$_2$. The fermionic system was then encoded into qubits using the JW transformation and ground state wavefunctions found through an offline VQE optimization. Finally, \mbox{ORDMs} were estimated from measurements on hardware allowing computation of information theoretic quantities such as entropy and mutual information.
A low-overhead noise reduction scheme was applied to mitigate noise from hardware errors and finite measurement shots, allowing us to get good agreement with noiseless simulations. 

In terms of the chemistry of the VC + $^1$O$_2 \rightarrow$ dioxetane reaction our results show that orbital correlations within the incoming O$_2$ are maximized in the intermediate stages of the reaction, as the O$_2$ molecule is stretched to accommodate the C-C bond before dioxetane is formed. 
However, when the local state of the O$_2$ molecule is fixed to a spin singlet (with total spin operator $\langle S^2 \rangle = 0$) we find all one-orbital correlations are entirely classical after accounting for SSRs ($E^{\text{SSR}}_i$), as observed previously for other systems~\cite{ding2020concept}. 
We also examine the $\langle S^2 \rangle = 2$ triplet spin configuration, which exhibits basis states with open shell occupations. From these results, we can see why open shell configurations of opposite spin are required for one-orbital quantum entanglement to contribute to the correlation of a chemical wavefunction (when the latter is dissected according to a given molecular orbital basis). This is due to the corresponding diagonal elements of the 1-ORDM being non-zero, which contribute to orbital entanglement.

Accounting for SSRs to maintain local fermionic symmetries not only allows for an operationally meaningful~\cite{ding2020concept} quantification of orbital correlations; by grouping the JW-transformed fermionic operators into mutually commuting sets, inclusion of SSRs additionally leads to significantly reduced numbers of measurement circuits (even more so than the significant reduction relative to classical evaluation~\cite{boguslawski13} without accounting for SSRs). 

In this work, measurements of ORDMs were performed using operator averaging of Pauli operator expectations, an approach that is suitable to current hardware.
Fault-tolerant quantum hardware would allow for alternative methods. For example, block encoding~\cite{rall20, steudtner23} of ORDM elements, amplitude estimation~\cite{knill07, simon24}, or approaches inspired by Lin and Tong's quantum phase estimation (QPE) method~\cite{lin22, zhang22spe}, each possibly combined with quantum error correcting schemes, to evaluate the ORDM element expectation values. Improvements on QPE-based techniques are also possible by using randomized circuit compilation \cite{zhiyan24}. Additionally, a recent gradient-based approach to estimate multiple expectation values could be used to capture all elements of an ORDM with an efficient number of queries to an oracle~\cite{huggins22}. While the scaling of these approaches can be asymptotically favorable compared to Pauli operator averaging, these methods typically have a large scaling prefactor, which prevents their application to current hardware due to large circuit requirements. 

Quantum computation offers novel opportunities for simulating chemistry from first principles, with the potential to overcome the complexity of the electronic many-body problem and tackle systems for which classical computation is prohibitively expensive~\cite{cao19, mcardle20, aryal23}. While orbital correlation and entanglement metrics have been computed classically in previous works, the demonstration of a quantum workflow to compute these metrics may enable future studies of larger chemical systems, allowing for the quantification of correlation and entanglement in regimes of chemistry beyond classical tractability.
The ability to represent chemical states with quantum circuits provides one motivation for use of a quantum computer in this work, since in principle a state can be stored on quantum hardware more efficiently (asymptotically) than classical approaches~\cite{peruzzo14, mcardle20}. 
We obtained chemical wavefunctions on a quantum computer using an offline optimized VQE algorithm. VQE has been used in many previous works on quantum computational chemistry~\cite{peruzzo14, mcclean16, mccaskey19, yordanov21, tilly22, sapova22, shee23, dalton24, guo24}, yet has well-known problems of large overhead in the number of measurements and susceptibility to exponentially decaying gradients. Other state preparation approaches without the drawbacks of VQE are possible, such as QPE-based methods~\cite{abrams99, zhao19}, and a recently published approach that reports optimal circuit depth~\cite{zhang22}. We also mention a new method which utilizes Lindblad dynamics to prepare ground states and requires only a single ancilla qubit, even when the initial approximation has vanishing overlap with the true ground state \cite{zding24_gs}. However, similar to the situation for measurement strategies outlined above, these methods have large scaling prefactors and are too demanding for current hardware.
Although the systems studied here are small enough to allow orbital entropies to be calculated classically, the approach we have demonstrated should scale well in terms of the number of measurable circuits with respect to qubits.
This approach will facilitate the quantification of the entanglement structure of larger and more complicated systems on future fault-tolerant hardware. Finally, we note that this work demonstrates the quantum computation of correlation and entanglement metrics on a molecular system undergoing strongly correlated transition states, which highlights the significance of this work in terms of novel applications of quantum computing to strongly correlated chemistry.

\section*{Data Availability}

\noindent The data supporting the conclusions of this paper are available from the corresponding authors upon request.

\section*{Code Availability}

\noindent The code used to produce the results and generate the figures of this paper are available from the corresponding authors upon request.

\section*{References}

\bibliography{references}

\begin{thebibliography}{84}
\bibitem [{ci_()}]{ci_coeffs}%
  \BibitemOpen
  \href@noop {} {}\bibinfo {note} {While CI coefficients obtained from VQE are close to the coefficients obtained from CASSCF, the latter involve small contributions from determinants other than the 4 chosen here, hence coefficients for these determinants recalculated from VQE are slightly different. Nevertheless, optimizing the VQE ansatz in the space of determinants restricted to those in Tables~\ref{tab:sv_coeffs_gates} and~\ref{tab:sv_coeffs_gates_triplet} recovers the CASSCF energy to approximately 10$^{-6}$ Ha or better.}\BibitemShut {Stop}%
\bibitem [{em_()}]{em_calcs}%
  \BibitemOpen
  \href@noop {} {}\bibinfo {note} {Noiseless emulator calculations were also performed (not shown) using 1,000, 10,000, and 100,000 shots per circuit, with the largest difference observed between 1,000 and 10,000 for the two-orbital entropies. At 100,000 shots per circuit, two-orbital entropies were closer to ideal values. However, as our results show, excellent agreement with ideal values can be obtained using 10,000 shots per circuit when combined with a low-overhead noise reduction scheme.}\BibitemShut {Stop}%
\end{thebibliography}

\section*{Acknowledgements}\label{sec:acknowledgements}

\noindent We thank David Zsolt Manrique and Kentaro Yamamoto for their feedback on an earlier version of this manuscript. We also acknowledge Chandan Kumar from BMW group and Elvira Shishenina for helpful discussions.

\section*{Author Contributions}

\noindent All authors designed and conceived the initial project. G.G.D. and C.N.S. implemented the methods, ran calculations and hardware measurements, and produced the figures. All authors contributed to technical discussions and writing of the manuscript. G.G.D. and C.N.S. contributed equally to this work.

\section*{Competing Interests}

\noindent The authors declare no competing interests.

\onecolumngrid
\newpage
\appendix

\setcounter{figure}{0}
\makeatletter
\renewcommand{\fnum@figure}{FIG. S\thefigure}
\makeatother

\section*{Supplementary Information for ``Measuring Correlation and Entanglement between Molecular Orbitals on a Trapped-Ion Quantum Computer''}

\section{Fermionic operators for the ORDMs}\label{app:ordm-fermion-ops}

\noindent Here we give the 16 elements of the 1-ORDM in terms of the fermionic operators of the molecular orbitals:
\begin{align}
\mathcal{O}(i)_1 &= 1 - \hat{n}_{i, \uparrow} - \hat{n}_{i, \downarrow} + \hat{n}_{i, \uparrow}\hat{n}_{i, \downarrow} \\
\mathcal{O}(i)_2 &= \hat{f}_{i, \downarrow} - \hat{n}_{i, \uparrow}\hat{f}_{i, \downarrow} \\
\mathcal{O}(i)_3 &= \hat{f}_{i, \uparrow} - \hat{n}_{i, \downarrow}\hat{f}_{i, \uparrow} \\
\mathcal{O}(i)_4 &= \hat{f}_{i, \downarrow}\hat{f}_{i, \uparrow} \\
\mathcal{O}(i)_5 &= \hat{f}^\dagger_{i, \downarrow} - \hat{n}_{i, \uparrow}\hat{f}^\dagger_{i, \downarrow} \\
\mathcal{O}(i)_6 &= \hat{n}_{i, \downarrow} - \hat{n}_{i, \uparrow}\hat{n}_{i, \downarrow} \\
\mathcal{O}(i)_7 &= \hat{f}^\dagger_{i, \downarrow}\hat{f}_{i, \uparrow} \\
\mathcal{O}(i)_8 &= \text{-}\hat{n}_{i, \downarrow}\hat{f}_{i, \uparrow} \\
\mathcal{O}(i)_9 &= \hat{f}^\dagger_{i, \uparrow} - \hat{n}_{i, \downarrow}\hat{f}^\dagger_{i, \uparrow} \\
\mathcal{O}(i)_{10} &= \hat{f}_{i, \downarrow}\hat{f}^\dagger_{i, \uparrow} \\
\mathcal{O}(i)_{11} &= \hat{n}_{i, \uparrow} - \hat{n}_{i, \uparrow}\hat{n}_{i, \downarrow} \\
\mathcal{O}(i)_{12} &= \hat{n}_{i, \uparrow}\hat{f}_{i, \downarrow} \\
\mathcal{O}(i)_{13} &= \hat{f}^\dagger_{i, \downarrow}\hat{f}^\dagger_{i, \uparrow} \\
\mathcal{O}(i)_{14} &= \text{-}\hat{n}_{i, \downarrow}\hat{f}^\dagger_{i, \uparrow} \\
\mathcal{O}(i)_{15} &= \hat{n}_{i, \uparrow}\hat{f}^\dagger_{i, \downarrow} \\
\mathcal{O}(i)_{16} &= \hat{n}_{i, \uparrow}\hat{n}_{i, \downarrow} ,
\end{align}
where $\hat{f}_{i, \sigma}$ and $\hat{n}_{i, \sigma}$ are the annihilation operator and number operator for spin orbital $\sigma$ of molecular orbital $i$.

\section{Qubit operators for the ORDMs}\label{app:ordm-qubit-ops}

\noindent We map the up and down modes of fermionic orbital $i$ to a pair of qubits $(2i, 2i+1)$ by JW transformation, using the convention
\begin{align}
&\hat{f}_{i,\uparrow} = \frac{1}{2} \bigg( X_{2i} + i Y_{2i} \bigg) \prod_{k > 2i} Z_k \\
&\hat{f}_{i,\downarrow} = \frac{1}{2} \bigg( X_{2i + 1} + i Y_{2i + 1} \bigg) \prod_{k > 2i + 1} Z_k .
\end{align}

The qubit operators for the diagonal elements of the 1-ORDM matrix are given by:
\begin{align}
\mathcal{O}(i)_1 &= \frac{1}{4} \bigg( \mathbb{I} + Z_{2i} + Z_{2i+1} + Z_{2i} Z_{2i+1} \bigg) \\
\mathcal{O}(i)_6 &= \frac{1}{4} \bigg( \mathbb{I} + Z_{2i} - Z_{2i+1} - Z_{2i} Z_{2i+1} \bigg) \\
\mathcal{O}(i)_{11} &= \frac{1}{4} \bigg( \mathbb{I} - Z_{2i} + Z_{2i+1} - Z_{2i} Z_{2i+1} \bigg) \\
\mathcal{O}(i)_{16} &= \frac{1}{4} \bigg( \mathbb{I} - Z_{2i} - Z_{2i+1} + Z_{2i} Z_{2i+1} \bigg).
\end{align}

Instead of giving the full expressions for the 2-ORDM qubit operators, we list the remaining $\mathcal{O}(i)_k$ needed to construct each of them
\begin{align}
\mathcal{O}(i)_2 &= \frac{1}{4} \bigg( X_{2i+1} + i Y_{2i+1} + Z_{2i} X_{2i+1} + i Z_{2i} Y_{2i+1} \bigg) \prod_{k > 2i + 1} Z_k \\
\mathcal{O}(i)_3 &= \frac{1}{4} \bigg( X_{2i} + i Y_{2i} + X_{2i} Z_{2i+1} + i Y_{2i} Z_{2i+1} \bigg) \prod_{k > 2i + 1} Z_k \\
\mathcal{O}(i)_{4} &= \frac{1}{4} \bigg( (1 - i) \, Y_{2i} Y_{2i + 1}  - (1 - i) \, X_{2i} X_{2i + 1} \bigg) \\
\mathcal{O}(i)_5 &= \frac{1}{4} \bigg( X_{2i+1} - i Y_{2i+1} + Z_{2i} X_{2i+1} - i Z_{2i} Y_{2i+1} \bigg) \prod_{k > 2i + 1} Z_k \\
\mathcal{O}(i)_{7} &= \frac{1}{4} \bigg( i Y_{2i} X_{2i+1} - i X_{2i} Y_{2i+1} + X_{2i} X_{2i+1} + Y_{2i} Y_{2i+1} \bigg) \\
\mathcal{O}(i)_{8} &= \frac{1}{4} \bigg( X_{2i} + i Y_{2i} - X_{2i} Z_{2i+1} - i Y_{2i} Z_{2i+1} \bigg) \prod_{k > 2i + 1} Z_k \\
\mathcal{O}(i)_9 &= \frac{1}{4} \bigg( X_{2i} - i Y_{2i} + X_{2i} Z_{2i+1} - i Y_{2i} Z_{2i+1} \bigg) \prod_{k > 2i + 1} Z_k \\
\mathcal{O}(i)_{10} &= \frac{1}{4} \bigg( i Y_{2i} X_{2i+1} - i X_{2i} Y_{2i+1} - X_{2i} X_{2i+1} - Y_{2i} Y_{2i+1} \bigg) \\
\mathcal{O}(i)_{12} &= \frac{1}{4} \bigg( X_{2i+1} + i Y_{2i+1} - Z_{2i} X_{2i+1} - i Z_{2i} Y_{2i+1} \bigg) \prod_{k > 2i + 1} Z_k \\
\mathcal{O}(i)_{13} &= \frac{1}{4} \bigg( (1 - i) \, X_{2i} X_{2i + 1}  - (1 + i) \, Y_{2i} Y_{2i + 1}  \bigg) \\
\mathcal{O}(i)_{14} &= \frac{1}{4} \bigg( X_{2i} - i Y_{2i} - X_{2i} Z_{2i+1} + i Y_{2i} Z_{2i+1} \bigg) \prod_{k > 2i + 1} Z_k \\
\mathcal{O}(i)_{15} &= \frac{1}{4} \bigg( X_{2i+1} - i Y_{2i+1} - Z_{2i} X_{2i+1} + i Z_{2i} Y_{2i+1} \bigg) \prod_{k > 2i + 1} Z_k .
\end{align}

\newpage

\section{Chemical State Circuits}\label{app:chem_circ}

\begin{figure*}[ht]
    \centering
    \includegraphics[width=\textwidth]{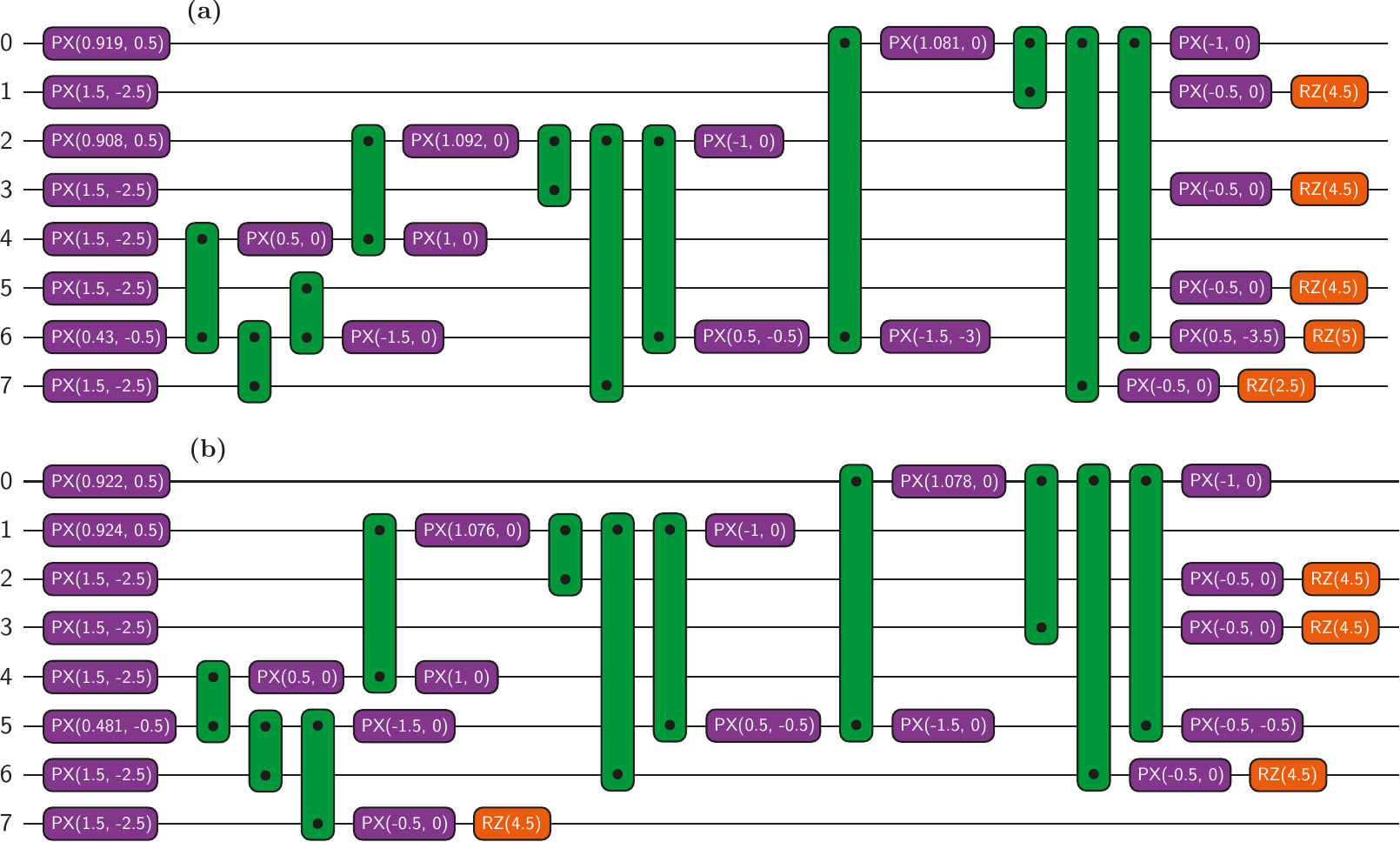}\
    \caption{Circuits to represent (a) the $\langle S^2 \rangle = 0$ wavefunction and (b) the $\langle S^2 \rangle = 2$ wavefunction, both corresponding to the 8 qubit O$_2$ $p$-orbital projected AVAS basis. Rotation angles (rounded to three decimal places) represent image 1 of the NEB path. The circuits are shown compiled to the H1-1 gate set~\cite{h11}. The circuits contain single qubit rotations about the $Z$ axes of the Bloch sphere (RZ), a more general single qubit rotation  called phased-$X$ that is equivalent to $\textrm{PX}(\theta, \phi) = R_Z(-\phi)R_X(\theta)R_Z(\phi)$ and a maximally entangling two-qubit rotation gate $e^{-i\frac{\pi}{4} Z \otimes Z}$. All rotation angles shown on the circuit are given in units of $\pi$.}
    \label{fig:chem_circ}
\end{figure*}

\noindent The optimized VQE circuits used to generate the ground state of image 1 of the NEB path in the singlet and triplet configurations are shown in Fig. S\ref{fig:chem_circ}. The statevectors of these circuits are equivalent to the second row, second column cell of (a) Table I and (b) Table II of the main text.

\section{Density correction to the hard-threshold for singular value}\label{app:threshold}

\noindent To reduce noise in our measured ORDMs we hard-threshold the singular values of the measured matrices. Due to the reduced number of elements in the matrices we measure we find we need to adjust the optimal threshold reported in~\cite{donoho2013optimal}. 

In~\cite{donoho2013optimal} the authors consider $d \times d$ matrices of the form $Y = X + \sigma Z$ where $X$ is the matrix whose singular values we want to find and $Z$ is a noise matrix whose elements are independent, identically distributed random variables of mean 0. They show that asymptotically the optimal hard-threshold for removing the small singular values of $Y$ caused by noise is $(4 / \sqrt{3}) \, \sqrt{d} \, \sigma$. 

\begin{figure*}[ht]
    \centering
    \includegraphics[width=\textwidth]{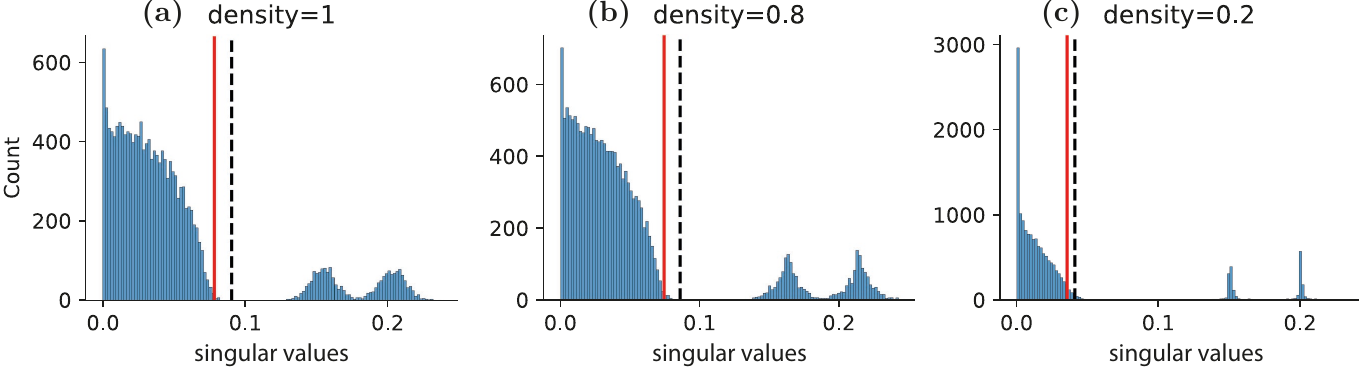}\
    \caption{
    Numerical investigation of the scaling of the ``bulk-edge'' of the singular values with the density of non-zero elements in a noisy matrix. For each density, $R$, we sample 1000 $16 \times 16$ matrices of the form $Y = X + \sigma Z$ where $X$ has all zero elements apart from $X_{1,1} = 0.15$ and $X_{2,2} = 0.2$ and $\sigma = 1/\sqrt{10000}$. The noise matrix $Z$ is constructed by randomly selecting $m = \lfloor R \times 16^2 \rfloor$ elements and setting each of those elements to values sampled from a Normal distribution with mean 0 and variance 1, all other entries of $Z$ are zero. Histograms for the singular values are shown at densities (a) $R=1$, (b) $R=0.8$ and (c) $R=0.2$. On each plot the red vertical line indicates an apparent ``bulk-edge'' of the noisy singular values at $2 \, \sqrt{d R} \, \sigma$ and the black dashed line is drawn at $(4 / \sqrt{3}) \, \sqrt{d R} \, \sigma$.
    }
    \label{fig:reduced-hardthreshold}
\end{figure*}

In characterizing the fermionic ORDM matrices we only measure $m$ elements of the matrix, so that $d^2 - m$ entries of $Y$ are zero. To compensate for this we find we should reduce the hard-threshold to $(4 / \sqrt{3}) \, \sqrt{d R} \, \sigma$, where $R = m / d^2$ is the density of non-zero elements in $Y$.
We numerically observe that this reduced hard-threshold is appropriate by plotting the distribution of singular values for 1000 randomly generated matrices at different densities, $R$, in Fig. S\ref{fig:reduced-hardthreshold}. We see that the ``bulk-edge'' of the noisy singular values decays with $\sqrt{R}$. Hence as the density of non-zero values in the matrix falls we should decrease the noise threshold too.

\end{document}